\documentclass[letterpaper,12pt]{article}
\pdfoutput=1 % if >0, submit to pdflatex for formatting

\usepackage{jheppub_X} % for details on the use of the package, please
\usepackage{booktabs,colortbl}
\usepackage{cancel,xcolor,multirow}
\usepackage[font=small]{caption}
\usepackage{subcaption}
\usepackage[capitalize]{cleveref}
  \crefname{table}{Table}{Tables}
  \crefname{section}{Sec.}{Secs.}
  \crefname{appendix}{App.}{Apps.}
\usepackage{comment}
\usepackage[normalem]{ulem}
\usepackage{tabu}
\usepackage{slashed}
\usepackage{xspace}
\usepackage{mathrsfs}
\usepackage{physics}
\usepackage[separate-uncertainty=true]{siunitx}
  \DeclareSIUnit{\pb}{\pico\barn}
  \DeclareSIUnit{\fb}{\femto\barn}
\usepackage{footnote}
\usepackage{adjustbox}
\usepackage{float}

\graphicspath{{./Plots/}}

\usepackage[T1]{fontenc}
\usepackage{listings}
\usepackage{lineno}
\providecommand{\ie}{\emph{i.e.}\xspace}
\providecommand{\eg}{\emph{e.g.}\xspace}

\providecommand{\ord}{O}

\DeclareRobustCommand{\Ref}[1]{Ref.~\cite{#1}}
\DeclareRobustCommand{\Refs}[1]{Refs.~\cite{#1}}

\providecommand{\Pythia}{\textsc{Pythia}\xspace}
\providecommand{\FastJet}{\textsc{FastJet}\xspace}
\providecommand{\Herwig}{\textsc{Herwig}\xspace}

\providecommand{\aS}{\ensuremath{\alpha_s}\xspace}

\newcommand{\ECF}{\ensuremath{e_2}\xspace}
\newcommand{\ECFb}{\ensuremath{e_2^{(\beta)}}\xspace}

\providecommand{\nfD}{\ensuremath{\widetilde{n}_F}\xspace}
\providecommand{\ncD}{\ensuremath{\widetilde{N}_C}\xspace}
\providecommand{\mD}{\ensuremath{\widetilde{m}_q}\xspace}
\providecommand{\aD}{\ensuremath{\widetilde{\alpha}}\xspace}
\providecommand{\LambdaD}{\ensuremath{\widetilde{\Lambda}}\xspace}

\DeclareMathOperator{\Li}{Li}

\newcommand{\be}{\begin{equation}} 
\newcommand{\ee}{\end{equation}} 
\newcommand{\bea}{\begin{eqnarray}}  
\newcommand{\eea}{\end{eqnarray}}
\newcommand{\bs}{\begin{split}} 
\newcommand{\es}{\end{split}}

\newcommand{\s}{\hspace{0.8pt}}

\definecolor{colorTC}{rgb}{.2,.7,.2}
\title{\Large 
Jet Substructure from Dark Sector Showers
}

\author[a]{Timothy Cohen,}
\author[a]{Joel Doss,}
\author[b]{and Marat Freytsis\,}

\emailAdd{tcohen@uoregon.edu}
\emailAdd{jdoss@uoregon.edu}
\emailAdd{marat.freytsis@rutgers.edu}

\affiliation[a]{\footnotesize Institute for Fundamental Science, Department of Physics, University of Oregon, Eugene, OR 97403}
\affiliation[b]{\footnotesize NHETC, Department of Physics and Astronomy, Rutgers University, Piscataway, NJ 08854}

\abstract{
We examine the robustness of collider phenomenology predictions for a dark sector scenario with QCD-like properties.
Pair production of dark quarks at the LHC can result in a wide variety of signatures, depending on the details of the new physics model.
A particularly challenging signal results when prompt production induces a parton shower that yields a high multiplicity of collimated dark hadrons with subsequent decays to Standard Model hadrons.
The final states contain jets whose substructure encodes their non-QCD origin.
This is a relatively subtle signature of strongly coupled beyond the Standard Model dynamics, and thus it is crucial that analyses incorporate systematic errors to account for the approximations that are being made when modeling the signal.
We estimate theoretical uncertainties for a canonical substructure observable designed to be sensitive to the gauge structure of the underlying object, the two-point energy correlator \ECFb, by computing envelopes between resummed analytic distributions and numerical results from \Pythia.
We explore the separability against the QCD background as the confinement scale, number of colors, number of flavors, and dark quark masses are varied.
Additionally, we investigate the uncertainties inherent to modeling dark sector hadronization.
Simple estimates are provided that quantify one's ability to distinguish these dark sector jets from the overwhelming QCD background.
Such a search would benefit from theory advances to improve the predictions, and the increase in statistics using the data to be collected at the high luminosity LHC.
}

\begin{document}
\maketitle
\flushbottom

%*********************************** Section ***********************************%
\setcounter{page}{2}
\section{Introduction}
\label{sec:intro}
%*******************************************************************************%
The physics program at the Large Hadron Collider (LHC) has reached a very mature stage.  
Run~II is now completed, and ATLAS and CMS each have $\sim \SI{150}{\per\fb}$ of \SI{13}{\TeV} data to explore.
This data has already taught us a variety of lessons regarding the Standard Model and beyond, but detection of new physics has thus far remained elusive.
Given the strong theory motivations provided by, \eg~supersymmetry and/or WIMP dark matter, most signal regions have been developed to target perturbative extensions of the Standard Model, which yield relatively clean, easily interpretable observables.
This is made sharp by the notion of Simplified Models~\cite{Alwall:2008va, Alwall:2008ag, Alves:2011wf}, which typically introduce one or two new physics states whose dynamics and interactions can be fully captured via a few additional terms that one adds to the Standard Model Lagrangian.
However, not all Standard Model extensions have collider signatures that can be captured in the weakly-coupled Simplified Model framework.
A good understanding of the novel signal regions associated with more out-of-the-box ideas is crucial to achieving full coverage when searching for new physics potentially being produced at the LHC.

Of particular relevance here is the idea that the dark matter could be a stable remnant of some new strong dynamics that resides in a hidden sector~\cite{Strassler:2006im, Han:2007ae, Harnik:2008ax, Strassler:2008fv, Juknevich:2009ji, Juknevich:2009gg, Carloni:2010tw, Carloni:2011kk, Seth:2011ci, Kribs:2016cew, Knapen:2016hky, Dienes:2016vei, Pierce:2017taw, Buchmueller:2017uqu, Kribs:2018oad, Kribs:2018ilo, Lee:2018pag, Beauchesne:2018myj, Alimena:2019zri, Bernreuther:2019pfb, Cheng:2019yai, Li:2019ulz, Brax:2019koq, Beauchesne:2019ato, Costantino:2020msc}.
It is then reasonable to assume the presence of some non-gravitational connection to the visible sector, such that the hidden sector was in thermal contact with the Standard Model at some point in the early Universe.  
This could result from a renormalizable interaction involving the Higgs, Neutrino, and/or Hypercharge Portals~\cite{Holdom:1985ag, Patt:2006fw, Falkowski:2009yz} or could be due to the exchange of some new mediator.
Depending on the properties of the portal, it could be possible to access the hidden sector at the LHC.
Furthermore, the dark strong dynamics could obfuscate the resultant signatures, as has been demonstrated concretely through many examples, \eg~lepton jets~\cite{ArkaniHamed:2008qp, Baumgart:2009tn, Chan:2011aa, Aad:2015sms, Buschmann:2015awa, Aad:2019tua}, emerging jets~\cite{Schwaller:2015gea, Sirunyan:2018njd, Renner:2018fhh}, semi-visible jets~\cite{Cohen:2015toa, Cohen:2017pzm, Park:2017rfb, Beauchesne:2017yhh}, and soft bombs~\cite{Harnik:2008ax, Knapen:2016hky}.

All of these examples share a common characteristic: a hard collision can generate a dark sector parton that subsequently undergoes a dark sector parton shower.
This often yields a high multiplicity of soft final state particles, smearing out the kinematics of the underlying partons and making it difficult to distinguish the associated signal against large backgrounds.
There is a further practical complication due to the fact that these signatures rely on the presence of dark strong dynamics --- the theoretical predictions are not nearly as well understood as in the Simplified Model case.
As a result, searches for this class of models are usually designed to be very inclusive, avoiding over reliance on details of the modeling.
The resulting trade-off between signal significance and systematic error mitigation motivates the work presented here: our goal is to understand the systematic uncertainties associated with making predictions that rely on dark sector strong dynamics.
An appreciation of which aspects of the observable can be reliably considered is crucial for the optimization of resulting search strategies.

Specifically, we focus on scenarios where the dark hadrons that result from a dark sector shower promptly decay back to Standard Model hadrons.
Our goal is to explore the properties of the resulting jets' substructure, and to quantify the uncertainty inherent to making such predictions.
Since substructure is sensitive to a variety of IR effects, such as the dark hadron mass spectrum and hadronization model, our work provides an observable-driven window into the systematic issues associated with making predictions for these strongly coupled dark sector scenarios.

As the use of jet substructure has become routine (see \Refs{Salam:2009jx, Abdesselam:2010pt, Altheimer:2012mn,  Altheimer:2013yza, Shelton:2013an, Adams:2015hiv, Cacciari:2015jwa, Larkoski:2017jix, Marzani:2019hun} for some reviews), many observables have been proposed to distinguish quark and gluons, or to tag boosted objects, and applications to dark sector showers have also been previously explored~\cite{Park:2017rfb}.
Detailed comparisons of parton and hadron level predictions for substructure observables have been performed in the context of the Standard Model, \eg~see the Les Houches 2017 report~\cite{Bendavid:2018nar}.
Of particular interest here are variables that were designed to be sensitive to the showering history of a jet, since our goal is to find ways to distinguish QCD jets from those that resulted from showering within a dark sector.
We are also interested in taking advantage of advancements in analytic calculations that rely on resummation techniques to capture the showering contribution to substructure.
To this end, our benchmark observable will be the energy correlation function \ECFb~\cite{Larkoski:2013eya}, where $\beta$ controls the sensitivity to wide-angle radiation; see \cref{eqn:e2beta} below for details.
We choose to focus on \ECFb since this family of observables is primarily sensitive to the gauge charge of the associated parton in the underlying hard process, which could be our only handle for uncovering dark shower signatures.\footnote{A number of observables has been considered for the problem for quark/gluon discrimination that are expected to provide superior discirminiation to \ECFb. These include both intrinsically IRC unsafe variables like track multiplicity or $N_{95}$~\cite{Pumplin:1991kc}, and also more complicated IRC safe observables that try to exploit correlations between multiple particles to approximate the behavior of these multiplicity variables~\cite{Moult:2016cvt,Frye:2017yrw,Larkoski:2019nwj}. The distributions of these observables are dominated by non-perturbative corrections, as discussed in more detail in \cref{sec:AnalTradResum}. For QCD, this information can be extracted from suitably chosen control regions, while in the case of a new hidden sector, our only recourse is to appeal to phenomenological models whose systematics are challenging to quantify. This implies that the ability to extract meaningful limits using such observables is significantly reduced, and so we will not consider them in this paper.}

There is potential concern when predicting the efficiency of jet-substructure assisted searches.
The discriminating power of nearly all substructure observables only becomes calculable if large logarithms that can appear in perturbation theory are resummed to all orders.
If this calculation is performed using a Monte Carlo generator such as \Pythia, only the leading logarithms (LL, defined in \cref{sec:frame}) are correctly captured, resulting in large expected theory uncertainties, which cannot be quantified by running the generator alone.\footnote{Automating parton showers beyond leading log and leading color is extremely challenging.  Some progress towards formalizing the problem was made in~\Ref{Nagy:2017ggp}, followed by a numerical approach to address aspects of subleading color in~\Ref{Nagy:2019pjp}.  
For recent progress in automating aspects of next-to-leading-logarithm accurate parton showers, see~\Refs{Alioli:2015toa, Hoeche:2017jsi, Forshaw:2020wrq}, with a recent candidate full proposal in~\Ref{Dasgupta:2020fwr}. }
For QCD studies, such concerns are partially ameliorated by the fact that the parameters of generators are tuned to real data, allowing them to often match the real world better than their formal accuracy would suggest. 
When looking for physics beyond the Standard Model that we have not yet observed, we have no such recourse.
To better address this state of affairs, we take advantage of theoretical technology developed to resum the soft and collinear QCD logarithms that contribute to \ECFb at leading and next-to-leading logarithmic order along with modern numerical implementations within \Pythia.
Sensibly enveloping across the spread of associated predictions will allow us to quantify the systematic error band that is the main result of this work.
These error bands can then be utilized to consistently include substructure information into LHC searches for dark sector physics.

Throughout this paper, we assume the dark sector includes \nfD families of dark quarks which bind into dark hadrons at energies below some dark confinement scale \LambdaD due to a non-Abelian dark $SU(\ncD)$ gauge group. 
Dark quarks will be produced with large transverse momentum $p_T \gg \LambdaD$ such that they shower and hadronize, yielding jets of dark hadrons. 
We assume that these dark hadrons decay promptly back to Standard Model quarks,\footnote{Decays to gluons are also in principle possible but to have them dominate the decay rate would require more involved model building.} yielding QCD-like jets.
We then explore the impact on the \ECFb observable as we vary the dark sector parameters \LambdaD, \nfD, \ncD, and the effect of making the dark quarks massive.
In addition, we provide an approximate characterization of the non-perturbative uncertainties associated with dark hadronization by exploring the impact of varying the phenomenological parameters associated with the Lund string model~\cite{Andersson:1983ia}.
We then use our error bars to estimate the extent to which dark sector showers can be distinguished from QCD when including the impact of substructure.

The rest of this paper is organized as follows.
In \cref{sec:frame}, we introduce the two-point energy correlation function, which will be used as our benchmark substructure variable.  
We then review how to calculate this observable to next-to-leading-logarithmic accuracy utilizing traditional resummation techniques. 
Our enveloping procedure that combines the analytic predictions with numerical results derived from \Pythia is then introduced, and provides a proxy for the systematic error associated with making a dark substructure prediction.
In \cref{sec:variations}, we present the extent to which the substructure changes as a function of some of the dark sector parameters: the dark confinement scale \LambdaD, the number of dark colors \ncD, dark flavors \nfD, and the dark quark mass \mD.
In \cref{sec:varyHad}, we explore the effect of varying the parameters that model the dark sector hadronization. 
In \cref{sec:discovery}, we estimate our ability to experimentally probe a dark sector jet against the QCD background. 
We present our conclusions in \cref{sec:end}. 
In \cref{sec:analytics}, we detail the expressions that are used to derive the analytic contributions to our systematic error envelopes.

%*********************************** Section ***********************************%
\section{Substructure Observables with Error Envelopes}
\label{sec:frame}
%*******************************************************************************%
A large array of jet substructure observables and algorithms have been developed, and are being combined in analyses in increasingly complicated ways.
However, the majority of substructure techniques are designed to find evidence of hard processes buried within boosted hadronic events,\footnote{For signals that yield high multiplicity final states via perturbative decays, so-called ``accidental substructure'' can also provide a useful handle, \eg~see~\cite{Hook:2012fd, Cohen:2012yc}.} and as such, most observables are optimized for the identification of distinct multi-prong structures within a jet.
A dark sector has no guarantee that it will produce such structure.
Instead, we are interested in observables that are sensitive to the structure of the color charge and gauge group of the radiation making up the parton shower.
This problem is closely analogous to the problem of quark/gluon discrimination in QCD, and we may look to prior work in this context for guidance~\cite{Gras:2017jty, Nilles:1980ys, Jones:1988ay, Fodor:1989ir, Jones:1990rz, Lonnblad:1990qp, Pumplin:1991kc, Gallicchio:2011xq, Gallicchio:2012ez, Krohn:2012fg, Chatrchyan:2012sn, Larkoski:2013eya, Larkoski:2014pca, Bhattacherjee:2015psa, FerreiradeLima:2016gcz, Bhattacherjee:2016bpy, Komiske:2016rsd, Davighi:2017hok, Metodiev:2018ftz, Larkoski:2019nwj}.
Additionally, we would like to work with infrared and collinear (IRC) safe observables, so that they are perturbatively calculable.
This is particularly important for a dark sector search since, unlike the situation for QCD, we have no data from which to extract any of the non-perturbative parameters which are required to make predictions.
Thus, there is no way to estimate their uncertainties without resorting to \emph{ad hoc} empirical models.

These two considerations almost uniquely limit us to considering observables which characterize the angular spread of radiation within the jet.
A representative choice is the two-point energy correlation function~\cite{Larkoski:2013eya}, defined as
\begin{equation}
\label{eqn:e2beta}
  \ECFb = \sum_{i<j \in J} z_i\s z_j\, (\theta_{ij})^\beta\,,
\end{equation}
where $\beta$ is the angular dependence parameter that determines how sensitive the variable is to the angular distribution of the radiation.
The jet algorithm determines the constituent particles in jet $J$ that are summed over in \cref{eqn:e2beta}.
In the context of a hadron collider like the LHC, it is most useful to define $z_i \equiv p_{T_i}/p_{T_J}$ and $\theta_{ij} \equiv R_{ij}/R_0$, where $p_{T_J}$ is the total $p_T$ of the jet, $R_{ij}$ is the Euclidean distance between the $i^\text{th}$ and $j^\text{th}$ partons in the $\eta$--$\phi$ plane, and $R_0$ is the jet radius.\footnote{For an $e^+ e^-$ collider, a more convenient choice would be $z_i \equiv E_i/E_J$ and $\theta_{ij} \equiv 2p_i \cdot p_j/E_i E_j$ or the actual Euclidean angle between the $i^\text{th}$ and $j^\text{th}$ partons. In the strict collinear limit, all these definitions collapse to be equivalent, and thus only differ in terms that are non-singular in the small \ECFb limit. We choose to normalize $\theta_{ij}$ by the jet radius $R_0$ to eliminate the leading dependence on $R_0$.}
For brevity, we will usually drop the $(\beta)$ superscript below when making general statements, and will also refer to the two-point energy correlation function as the energy correlator when appropriate from context.
Note that $e_2^{(\beta)}$ is equivalent to the $C_1^{(\beta)}$ variable introduced in \Ref{Larkoski:2013eya} and widely used in experimental studies~\cite{Aaboud:2019aii, Sirunyan:2018asm, ATLAS:2016wzt}.

To build intuition, one can consider a jet with two constituents; in the infrared and collinear limit, the jet mass is given by $m^2/p_T^2 \simeq z_1 z_2 \bigl(\theta_{12}/R_0\bigr)^2$, such that $e_2^{(2)} \simeq m^2/p_T^2$.  
Hence, \ECFb can be seen as a generalization of jet mass that incorporates arbitrary angular dependence.
It is also closely related to the family of jet angularities~\cite{Berger:2003iw,Almeida:2008yp}, without the need to define a jet axis.

Our essential idea is to calculate the distributions of interest analytically and numerically assuming various approximations, and then use these to determine an error bar such that it spans the range of predictions.
First, we review the analytic calculation of the resummed substructure distributions at leading and next-to-leading log order, followed by a brief discussion of the numerical implementation using \Pythia.
Then, we explain how we combine the various approximations into an error envelope in the context of a QCD calculation.
This will set the stage for \cref{sec:variations}, where we explore the range of predictions for the substructure distributions resulting from a dark sector shower.

%**************** Subsection *******************************
\subsection{Analytics Using Traditional Resummation Techniques}
\label{sec:AnalTradResum}
%*************************************************************
To understand the robustness of the \ECF distributions, it is useful to explore the range of predictions that result from analytic techniques for calculating the normalized differential cross section.
These formulas were derived in \Ref{Larkoski:2014wba}, and we present a summary of the main steps for the calculations in \cref{sec:analytics}.
The collinear limit of the leading order $\ECF$ distribution generates a collinear logarithm from the integral over the splitting angle $\theta$ and a soft logarithm from the integral over the momentum fraction $z$.
Enforcing the kinematics of two-body momentum conservation with a delta function, we can write down the differential distribution for \ECF by appealing to the definition in~\cref{eqn:e2beta}:
\begin{equation}
\label{eqn:sigmaLO}
  \frac{1}{\sigma} \dv{\sigma_i^\text{LO}}{\ECF} =
    \frac{\aS}{\pi} \int_0^{R_0} \frac{\dd{\theta}}{\theta} \int_0^1 \dd{z}
      p_i(z)\, \delta\Biggl( z(1-z) \biggl( \frac{\theta}{R_0} \biggr)^\beta - \ECF \Biggr)\,,
\end{equation}
where $R_0$ is the jet radius\footnote{Although we work in a small jet radius limit, this is known to be a reasonable approximation even up to $R_0 \sim 1$~\cite{Dasgupta:2007wa,Dasgupta:2012hg}.} and $p_i(z)$ is the appropriate parton splitting function for a quark-initiated jet or a gluon-initiated jet, which are given by
\begin{align}
\label{eqn:splittings}
  p_q(z) & = P_{g \leftarrow q}(z) = C_F\frac{1+z^2}{1-z}\,, \notag\\
  p_g(z) & = \frac{1}{2}P_{g \leftarrow g}(z)+n_F P_{q \leftarrow g}(z) \notag\\
           & = C_A\biggl(\frac{z}{1-z}+\frac{1-z}{z}+z(1-z)\biggr)+n_F T_R\bigl(z^2+(1-z)^2\bigr)\,,
\end{align}
where $T_R = \frac{1}{2}$ is the index of the quark representation, \ie, the fundamental representation.
These splitting functions encode the divergences associated with a shower that is initiated by the emission of a soft gluon.

In the limit where $\ECF \ll 1$, we can simplify $z(1-z)(\theta/R_0)^{\beta} \simeq z(\theta/R_0)^{\beta}$ by assuming $z \ll 1$.
It is then straightforward to evaluate \cref{eqn:sigmaLO}, which yields
\begin{equation}
\label{eqn:sigmaLO_approx}
  \frac{\ECF}{\sigma} \dv{\sigma_i^\text{LO}}{\ECF} \simeq
    \frac{2\aS}{\pi} \frac{C_i}{\beta}
      \Biggl( \ln\frac{1}{\ECF} + B_i + \ord\bigl( \ECF \bigr) \Biggr)\,,
\end{equation}
where $C_q = C_F = \frac{N_C^2-1}{2N_C}$ and $C_g = C_A = N_C$ are the color factors associated with the jet, and $B_q = -\frac{3}{4}$ and $B_g = -\frac{11}{12} + \frac{n_F T_R}{3C_A}$ encode the subleading terms in the splitting functions that arise from hard collinear emissions. 
Identifying the characteristic logarithm $L \equiv \ln\big(1/\ECF\big)$, the cumulative distribution at leading order exhibits a characteristic double logarithm in the limit of small \ECF: 
\begin{equation}
\label{eqn:cumLO_approx}
  \Sigma_i^\text{LO} \equiv
    \int_0^{\ECF}\dd{x} \frac{1}{\sigma} \dv{\sigma_i^\text{LO}}{x}  =
      1 - \frac{\aS}{\pi} \frac{C_i}{\beta} \Bigl(L^2+2B_i L + \ord(L^0) \Bigr)\,.
\end{equation}
This shows that perturbation theory breaks down in the limit of small \ECF, so we would like to resum this double logarithm to derive a convergent prediction.

The authors of \Refs{Banfi:2004yd,Banfi:2004nk} derived a concise expression for the next-to-leading logarithmic (NLL) resummation of the cumulative distribution for recursively IRC safe observables such as \ECF:
\begin{equation}
\label{eqn:cumNLL}
  \Sigma_i^\text{NLL} =
    \frac{e^{-\gamma_E R_i'}}{\Gamma(1+R_i')}
      e^{-R_i} e^{-\frac{\alpha_s}{\pi} (R_{1,i} - G_{2,i} L^2 - G_{1,i} L)},
\end{equation}
where the ``radiator'' $R_i$ is given by
\begin{equation}
\label{eqn:radiator}
  R_i = \int_0^{R_0} \frac{\dd{\theta}}{\theta} \int_0^1 \dd{z} p_i(z) \frac{\aS(\kappa)}{\pi}
          \Theta\Biggl( z\biggl(\frac{\theta}{R_0}\biggr)^\beta - \ECF \Biggr),
\end{equation}
with $R'_i \equiv \dv{R_i}{L}$ and $\kappa = z \theta p_{T_J}$, and $R_{1,i}$ is the fixed-order (FO) correction at next-to-leading order, which allows one to match (in the Log-$R$ scheme~\cite{Catani:1992ua}) between the resummed and perturbative regimes, ensuring the appropriate kinematic endpoint is respected.
As such, $G_{2,i} L^2$ and $G_{1,i}L$ are the logarithms appearing in the fixed-order expression (in the collinear limit) that must to be subtracted to avoid double counting the resummed logarithms.
Simplifying $z(1-z)$ to $z$ in \cref{eqn:sigmaLO} is justified by the identical structure of the two collinear limits and is compensated by a suitable combinatoric factor, as further discussed in \cref{sec:analytics}.

In the context of quark/gluon discrimination, a number of observables have been proposed that seemingly satisfy our property of being perturbatively calculable while claiming to offer improved discrimination over the energy correlation function above~\cite{Moult:2016cvt,Frye:2017yrw,Larkoski:2019nwj}.
This comes at a price.
Instead of contributions from individual emissions contributing linearly to the observable, each emission's weight depends on the entire shower history.
However, this feature also increases the resulting sensitivity to non-perturbative corrections by reducing the parametric suppression of these effects, and until more detailed understanding of these features is available, it is difficult to recommend the use of such substructure variables in situations where these effects cannot be constrained by data.
Note that even in the case of the better understood \ECFb, the $\beta$ dependence of quark/gluon discrimination has been measured, and it noticeably deviates from that of the perturbative predictions~\cite{Aad:2014gea}.

An analytic evaluation of $R_i$ is possible, although challenging, \eg~see~\Ref{Marzani:2017mva}.
The calculation of the resulting efficiencies at NLL due to a cut on \ECF requires evaluating the gauge coupling \aS at two-loop order using the CMW scheme~\cite{Catani:1990rr}, such that efficiencies still need to be computed numerically.
Another issue is related to \aS becoming non-perturbative as the integral is evaluated at low enough scales.
To mitigate these complications, we follow the procedure outlined in \Ref{Larkoski:2014wba}: the coupling is only run at one-loop order and is frozen at a ``non-perturbative scale'' $\mu_\text{NP} = 7\Lambda$, where the factor of 7 is an arbitrary choice.
This allows us to find a closed-form solution to \cref{eqn:radiator} at the expense of limiting its logarithmic accuracy.
We will call this approximate evaluation of \cref{eqn:cumNLL} the ``modified leading logarithmic'' (MLL) resummed cumulative distribution with FO corrections.
All analytic distributions presented will be MLL+FO accuracy (with the exception of \cref{fig:QCDenvelope}).

%**************** Subsection *******************************
\subsection{Numerics From \Pythia}
%*************************************************************
Our analytic expressions have the benefit that they are transparent, in that we can precisely identify the approximations that go into the calculations.
However, they do not account for important corrections from, \eg~hadronization or finite quark masses.
They also do not provide any way for us to assess the impact of dark sector hadronization on our prediction.
To address these shortcomings, we compare our results for the \ECF observables to those of a Monte Carlo parton shower that models a new confining gauge group.
Although all parton showers in common use are formally accurate to leading log, they include various corrections with the goal of modeling certain higher order effects, \eg~see the Monte Carlo Event Generators review in~\Ref{Tanabashi:2018oca}.
It is worth emphasizing here that all such corrections assume QCD, and as such should be revisited in the context of more general confining theories.
Specifically here, we simulate events using \Pythia 8.240~\cite{Sjostrand:2014zea}.
We simulate $p\s p$ collisions at $\sqrt{s} = \SI{14}{\TeV}$ including initial- and final-state radiation (without multiple parton interactions) for all our events. 
The signal is generated via a direct portal from $\bar{q}\s q$ pairs to dark sector quarks, and the evolution of the dark sector is implemented in \Pythia's Hidden Valley module~\cite{Carloni:2010tw, Carloni:2011kk, Seth:2011ci}, including a dark parton shower, hadronization, and decay back to Standard Model states.
Events are clustered into anti-$k_t$ jets~\cite{Cacciari:2008gp} with radius $R_0 = 1.0$ and \ECF computed for each jet using \FastJet 3.3.2~\cite{Cacciari:2011ma}, subject to a jet-level cut of $p_T > \SI{1}{\TeV}$.

We will briefly comment on the implementation of the parton shower in \Pythia's Hidden Valley module.\footnote{For a more complete discussion, see the corresponding section of the \Pythia manual:\\ \url{http://home.thep.lu.se/~torbjorn/pythia82html/HiddenValleyProcesses.html}.} 
The underlying physics model is the same as that used for the time-like QCD shower.
Showering proceeds via the emission of a dark gluons from both dark quarks and gluons.
The dark quarks may be duplicated up to eight flavors, \nfD, with identical masses and integer spin by default (we set the dark quark spin to be $1/2$ in this paper).
Running of the dark gauge coupling is included up to one-loop for an arbitrary $SU(\ncD)$ gauge group, assuming massless quarks.
Although the functionality to include an arbitrary dark quark mass spectrum is available, we take the masses \mD to be degenerate throughout this study. 
We do not include any states that are charged under both the Standard Model and dark sector symmetry groups, although such states may be considered to extend the range of phenomenological handles in the resulting signal.

A number of aspects of our analytic calculation make its perturbative accuracy greater than that of the \Pythia parton shower.
Dark gluon splitting into quark pairs is not currently implemented in \Pythia; the $P_{q \leftarrow g}(z)$ splitting function is not singular in the soft limit, and therefore provides contributions beyond LL accuracy.
A choice of minimum allowed $p_T$ for emissions controls the termination of the shower at low scales.
This threshold may be tuned to data in the case of QCD, but for a dark parton shower, this is a parameter that should not be much larger than the confinement scale.
Matrix element corrections ensuring the accuracy of parton splitting to one-loop order are included in the QCD parton shower of \Pythia but, being model-dependent, not for the Hidden Valley module.
Comparing the analytic results to the \Pythia predictions will estimate the resulting uncertainties, which are either included or (in the case of the $p_T$ cut) have no impact on our analytic results.

The dark sector is assumed to confine, and hadronization is implemented via the Lund string model~\cite{Andersson:1983ia}, which has some associated parameters whose values are unknown a priori.
We explore the consequences of this fact below in \cref{sec:varyHad}.
Hadronization proceeds exclusively to dark pions and dark rho mesons, which all decay back to the Standard Model using flat matrix elements (assuming no flavor symmetries leading to stable dark mesons, see \eg \Refs{Cohen:2017pzm,Beauchesne:2017yhh}).
\Cref{tab:hiddenparams} enumerates the relevant parameters discussed along with their default settings.

%\clearpage
\begin{savenotes}
\begin{table}[t!]
  \renewcommand{\arraystretch}{1.6}
  \setlength{\arrayrulewidth}{.3mm}
  \centering
  \setlength{\tabcolsep}{0.6 em}
  \makebox[\linewidth]{\begin{tabular}{l|c|l|c|l|c}
  \multicolumn{2}{c}{Model} & \multicolumn{2}{|c|}{Showering} &
    \multicolumn{2}{c}{Hadronization} \\ \hline\hline
  \texttt{Ngauge}   & 3   & \texttt{FSR}        & on             &
    \texttt{fragment}   & on   \\ \hline
  \texttt{nFlav}    & 5   & \texttt{alphaFSR}   & ---            &
    \texttt{probVector} & 0.75 \\ \hline
  \texttt{spinFv}\footnote{\texttt{spinFv} controls the spin of particles charged under both the Standard Model and dark sector. If this flag is nonzero (zero), then the dark quark spins are forced to be either 0 or 1 ($1/2$).}           & 0   & \texttt{alphaOrder} & 1              &
    \texttt{aLund}      & 0.3  \\ \hline
  \texttt{spinqv}   & --- & \texttt{Lambda}     & \SI{1}{\GeV}   &
    \texttt{bmqv2}      & 0.8  \\ \hline
  \texttt{doKinMix} & off & \texttt{pTminFSR}   & \SI{1.1}{\GeV} &
    \texttt{rFactqv}    & 1.0 \\
\end{tabular}}
\caption{List of variables within the \Pythia 8.240 Hidden Valley module, along with the default choices made for the study performed here. All of these variables should be prepended with  ``\texttt{HiddenValley:}'' when being called within \Pythia.  Note that \texttt{spinqv} (\texttt{alphaFSR}) are derived from \texttt{spinFv} (\texttt{Lambda}), which is why they are marked with ``---'' in the table. Decay tables for the dark mesons must additionally be specified.}
\label{tab:hiddenparams}
\end{table}
\end{savenotes}

\clearpage

\subsection{Error Envelopes}
\begin{figure}[t!]
\center
\includegraphics[width=\linewidth]{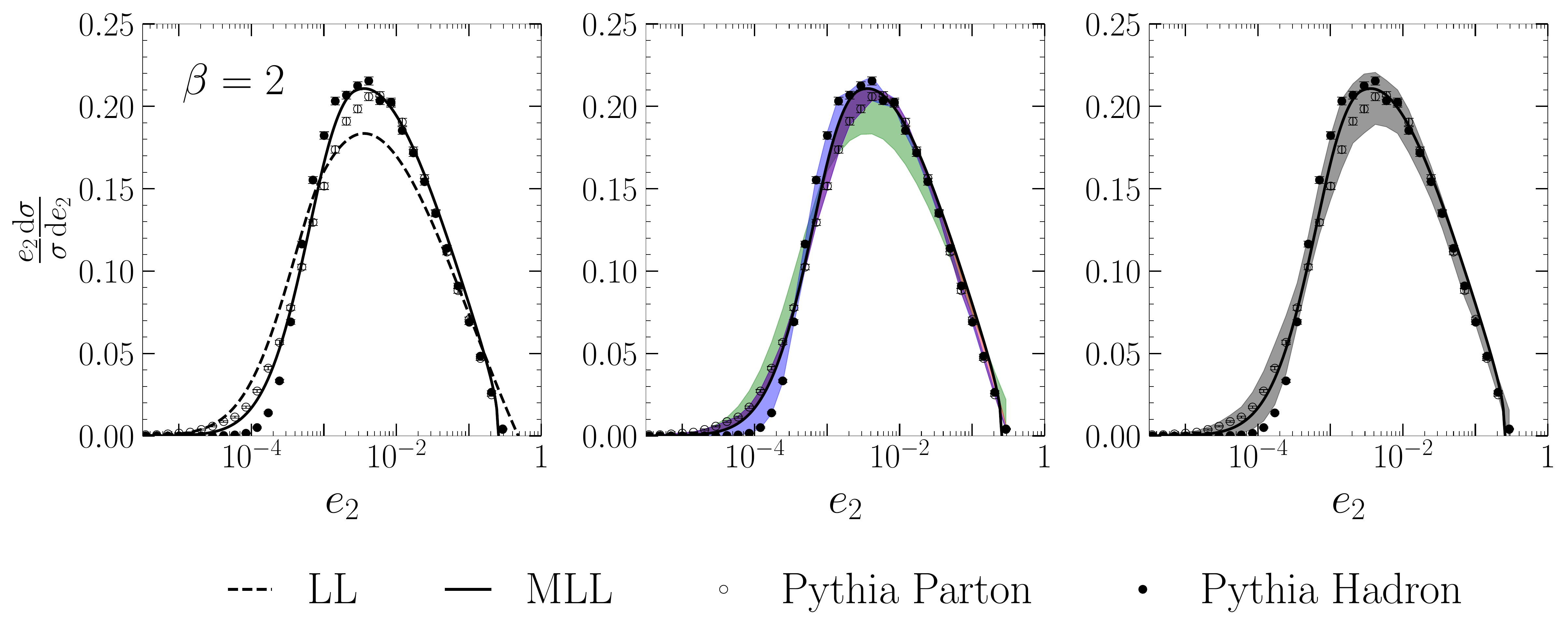}
\caption[The LOF caption]{Illustration of the enveloping procedure utilized to estimate theory systematic errors in this paper as applied to QCD, for $p_T = \SI{1}{\TeV}$ jets.
The analytic and numerical predictions for the \ECF distributions are shown assuming various levels of approximation as detailed in the legend [left panel].\footnotemark\ 
These predictions are used to create error sub-envelopes [middle panel] that are then combined to result in the final envelope [right panel].
The green sub-envelope captures the difference between the LL and MLL analytic predictions
The red sub-envelope captures the difference between the MLL analytic prediction and the parton level numerical result from \Pythia.
The blue sub-envelope captures the difference between the parton and hadron level prediction from \Pythia.
These envelopes are added in quadrature to compute the total envelope [right panel].
Note that when computing envelopes for the dark sector, we include three numerical predictions when the final states are dark partons, dark hadrons, or visible hadrons.
The angular dependence parameter is set to $\beta = 2$ for illustration.
}
\label{fig:QCDenvelope}
\end{figure}

In this section, we describe the procedure used to compute the error envelopes presented in \cref{sec:variations}.
To capture the ``perturbative'' theoretical uncertainty associated with these distributions, we combine a number of variations that probe the systematic uncertainties inherent to making dark shower predictions.
First, to incorporate uncertainties in the showering step, we capture the range of parton level predictions by comparing the LL order and the MLL~+~FO order analytics (which we refer to as MLL in the figures).
Next, we compare the MLL order analytics and the parton level numerics, \ie, turning off hadronization.  
Finally, we compare the parton level and the hadron level numerics to account for the effects of hadronization. 
For events originating from a dark sector shower, we also compare the dark hadron level and the visible hadron level numerics to capture the effects of decaying dark hadrons and their subsequent recombination into Standard Model hadrons. 
To construct our error bands, we sum the widths of these comparison sub-envelopes in quadrature to produce an averaged final envelope.
The results of this procedure when applied to QCD are presented in \cref{fig:QCDenvelope}.
Note that for the later plots we show the central value of the envelope merely to guide the eye; this curve does not simply follow from our analytic results.
Then in \cref{sec:Had} below, we investigate the uncertainty due to hadronization modeling.
The total error band that includes the perturbative and hadronization errors is then used as the input to our search sensitivity estimates for the LHC presented in \cref{sec:discovery}.

\footnotetext{Note that the excellent agreement between the analytic and numerical distributions here does not persist across parameter variations (and may actually be due to the fact that \Pythia is tuned using jet mass as one of the inputs).}
We note that a common approach to calculating a theory uncertainty is to vary factorization, resummation, and (when considering exclusive observables) fragmentation scale parameters by a factor of two away from their canonical choices.  
This is a way of estimating higher-order terms that have not been explicitly computed by assuming they are dominated by their logarithmically enhanced pieces.
The logarithms dominating our distributions are not due to a running effect so that uncertainties in the resummation procedure will not be captured by such an approach.
Theoretical uncertainties for resummed calculations typically require more involved multi-scale variational schemes using effective field theory frameworks.\footnote{For work on adopting such variations in traditional resummation techniques, see \Ref{Almeida:2014uva}. Alternative schemes for estimating theory errors have also been introduced in the context of Standard Model calculations, \eg~see~\Ref{TackmannSCET}.}

The enveloping approach advocated here is designed to incorporate this uncertainty, while also accounting for unknown details of the hadronization and decay properties of the dark sector.
%
%As such, our error envelopes are a conservative estimate of the theory uncertainty for substructure modeling throughout the dominant part of our distributions.
%
Our estimates of perturbative errors are comparable to those of the effective field theory scale variational approaches for QCD, where similar calculations have been done~\cite{Frye:2016aiz}.
Depending on the precise treatment of the normalization when taking scale variations, it is possible to find significantly larger errors below the Sudakov peak (\eg~see Fig.\ 5 in \Ref{Jouttenus:2013hs}), where the interplay of constraints from the integrated cross section calculation and breakdown of resummation convergence makes uncertainties particularly sensitive to choice of scheme~\cite{Bertolini:2017eui}.
However, the resulting effect on signal yields is minimal, since such large uncertainties occur in a vary rapidly falling part of the distribution.

Before showing the results from varying parameters in the dark sector, we note that the analytic approximation for the radiator $R_i$ used in our calculations is not continuously differentiable, see \cref{eqn:radiator_g1_A,eqn:radiator_l1_A,eqn:radiator_e1_A}.
This is a consequence of sharply cutting off the integrals using the non-perturbative scale $\mu_\text{NP}$ introduced in \cref{sec:AnalTradResum} above, which leads to a kink in the second derivative of the radiator $R''_i$. 
To avoid this issue, we follow \Ref{Larkoski:2014wba} and replace this derivative with a discrete approximation:
\begin{equation}
\label{eqn:logderiv}
  R''_i \bigl(\ECF\bigr) \simeq
  \frac{R'_i\bigl(e^{-\delta} \ECF\bigr) - R'_i(\ECF)}{\delta}\,,
\end{equation}
where the choice $\delta = 1$ is an additional source of theoretical uncertainty that is negligible to single logarithmic accuracy.

\begin{figure}[t!]
\center
\includegraphics[width=.5\linewidth]{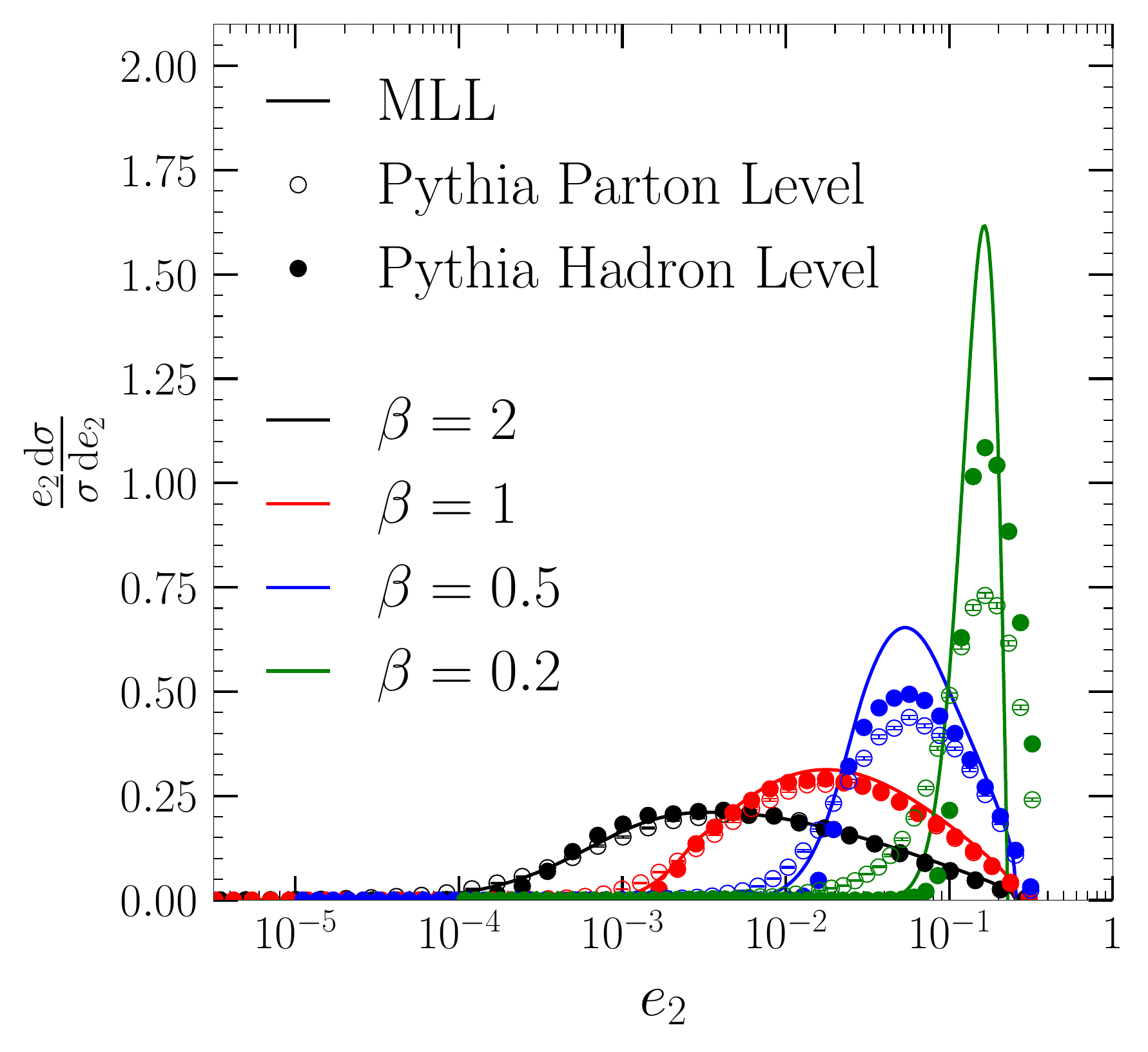}
\caption{Dependence on the angular dependence parameter $\beta$ in QCD for $p_T = \SI{1}{\TeV}$ jets.
We show the predictions derived using the MLL analytic calculation, along with the parton and hadron level numerical results from \Pythia.
Larger (smaller) angular dependence emphasizes the contribution from pairs of partons with larger (smaller) angular distance.
The analytic calculations begin to break down for angular dependence values $\beta < 0.5$, which is reflected here in the fact that the $\beta = 0.2$ curve does not appropriately terminate at the kinematic endpoint.}
\label{fig:C1QCD}
\end{figure}

\Cref{fig:C1QCD} shows the analytic and numeric $\ECF$ distributions for QCD jets across various angular dependence values $\beta$. 
We note the agreement between the analytic and the numeric distributions begin to diverge for low angular dependences $\beta < 0.5$. 
Furthermore, the $\beta = 0.2$ analytic distribution does not appropriately terminate at the kinematic endpoint. 
We conclude that even though we are working in parameter space where the resummation techniques should be a good approximation, the low angular dependence regime of $\ECF$ is not well modeled. 
For this reason, we will focus our analysis on the behavior of the $e_2^{(2)}$ and $e_2^{(0.5)}$ to explore the impact of varying $\beta$.

From the definition of \ECF in \cref{eqn:e2beta}, we see that increasing $\beta$ gives greater weight to emissions at larger angular distances in the distribution.
Since emissions at large angle within a jet are preferentially softer at large angles, giving lower weight to large-angle emissions leads to \ECF distributions closer to their kinematic endpoint, behavior that is clearly reflected in \cref{fig:C1QCD}.
Simultaneously, the distribution of \ECF is dominated by emissions in singular regions of phase space, so that lower values of $\beta$ provide more sensitivity to the structure of the collinear singularity of partonic splitting functions.
This comes at the cost of loss of perturbative control.
\Cref{sec:AnalTradResum} makes clear that the effective coupling in the calculation of \ECFb is $\aS/\beta$ and that for values of $\beta \ll 1$, perturbative control of the \ECF distribution is lost throughout phase space.

\subsection{Applying the Predictions}
\label{sec:Apply}
In addition to plotting the normalized \ECF distributions, we will provide a few different ways of presenting the predictions.
We will show the cumulative cross section, which is derived by taking the differential distribution and numerically evaluating the following integral, see the bottom row of Figs.~\ref{fig:C1Lambda}, \ref{fig:C1Color}, and \ref{fig:C1Flavor}:
\begin{equation}
\label{eqn:xcut_Error}
  \Sigma(x_\text{cut}) =
    \int_{x_\text{cut}}^{e_{2,\max}} \dd{\ECF} \frac{1}{\sigma} \dv{\sigma}{\ECF}\,,
\end{equation}
where $e_{2,\max} = \frac{1}{4}R_0^{\beta}$.
To incorporate the error envelopes, we assume they are fully correlated.
In practice, this simply means we compute the upper (lower) error envelope of the cumulative distribution by integrating the upper (lower) edge of the differential distribution.
The choice of $x_\text{cut}$ will be optimized below when we discuss the discovery potential of dark substructure in \cref{sec:discovery}.

We also provide some quantitative insight into how different the signal and background distributions are using the MLL analytic predictions directly, see Figs.~\ref{fig:C1LambdaDisc}, \ref{fig:C1ColorDisc}, and \ref{fig:C1FlavorDisc}.
The left and middle panels of these figures provide two different figures of merit, which give a quantitative sense of how well one could distinguish signal from background, in this case approximated by quark initiated jets.
Specifically, on the left we show ROC curves, which are the parametric curve that traces the background rejection $1-\epsilon_B$ as a function of signal acceptance $\epsilon_S$, due to varying a cut on \ECF.
The middle panels show the parametric curve for discovery significance $\epsilon_S/\sqrt{\epsilon_B}$ as a function of signal acceptance $\epsilon_S$, again due to varying a cut on \ECF.
The right panels show the change in the signal rate as a function of the dark sector parameter that is being varied, for a benchmark fixed background factor, which is taken to be $1-\epsilon_B = 90\%$.
As we will explore in the next section, these various ways of presenting the predictions provide additional insight into the behavior of the \ECF observable across the dark sector parameter space.

%*********************************** Section ***********************************%
\section{Distinguishing Dark Substructure from QCD}
\label{sec:variations}
%*******************************************************************************%
Now that we have established a method to estimate the theoretical uncertainties inherent to calculating substructure distributions, we will apply this technology to explore the range of predictions one can expect from a dark sector including error bars.
This demonstrates the behavior of the dark sector as a function of its parameters. 
In particular, we will highlight how the uncertainties depend on the parameters.
While we incorporate the effects of hadronization in this section, we set the hadronization parameters to their default values.
The results presented here will be combined with an estimate of hadronic uncertainties in \cref{sec:Had}, which are computed by varying the non-perturbative parameters.
These are then used as the inputs to the estimates performed in \cref{sec:discovery}, where we study to what extent it is possible to distinguish dark sector showers from QCD via substructure measurements. 
Note that we have made the simplifying assumption that the QCD background is entirely composed of quark jets in what follows, since the signal will dominate in the central region of the detector.
A more realistic study should of course incorporate a more sophisticated modeling of the background.
However, since our uncertainties are dominated by the signal modeling, a more careful accounting of the quark/gluon composition of the background should be a subdominant effect.

%**************** Subsection *******************************
\subsection{\boldmath $\LambdaD$ Dependence}
%*************************************************************

\begin{figure}[t!]
\center
\includegraphics[width=\linewidth]{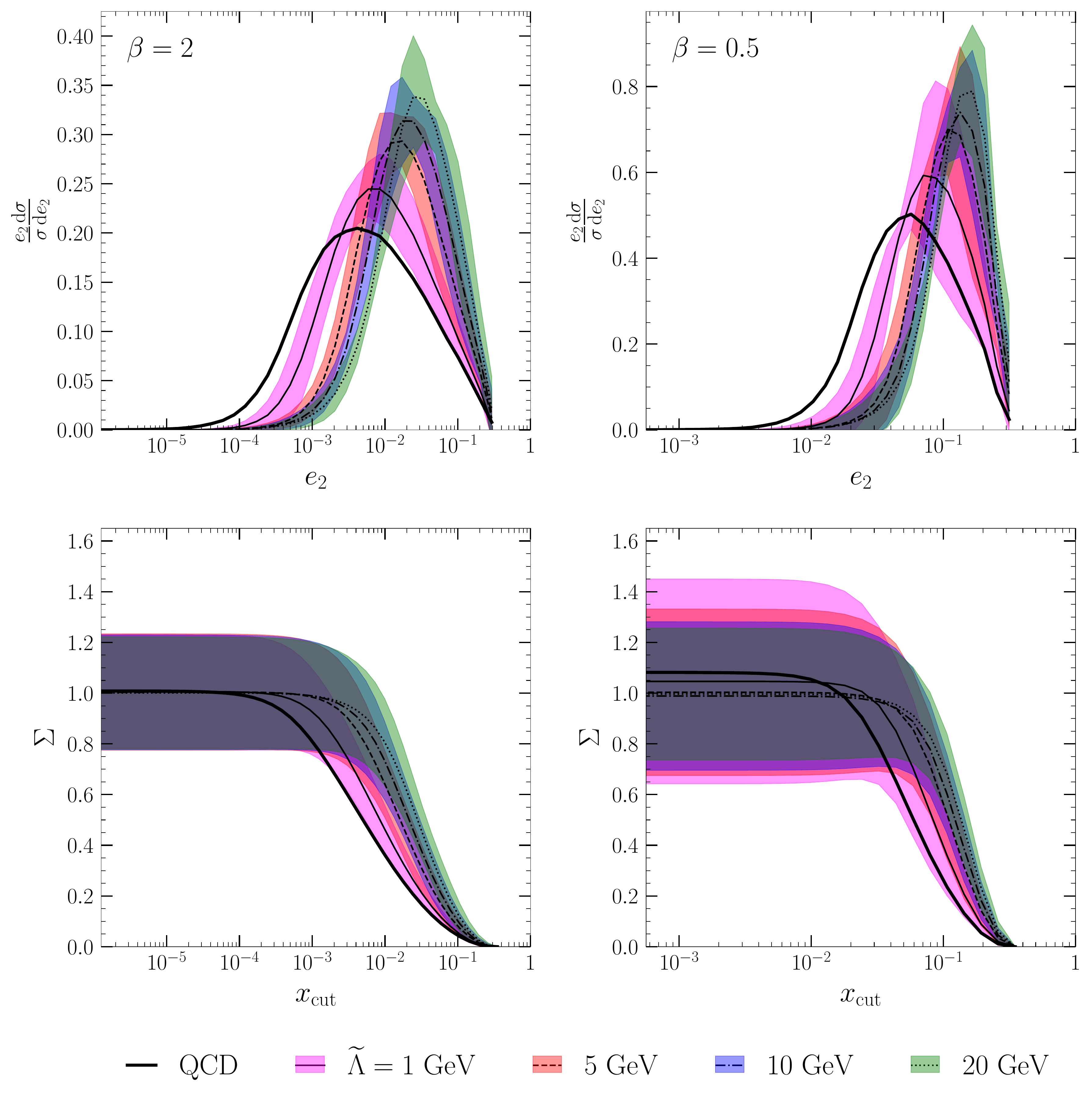}
\caption{The behavior of \ECF as the dark confinement scale \LambdaD is varied, for $p_T = \SI{1}{\TeV}$ jets.
See the legend for values of \LambdaD; all other values are given in \cref{tab:hiddenparams}.
We show the normalized \ECF distributions [top], where the central value of the envelope is marked with the black lines, while the shaded region denotes the envelope.
The peak shifts to larger values of \ECF as \LambdaD is increased.
The cumulant distributions $\Sigma$ as a function of $x_\text{cut}$ are also provided [bottom], where again the lines denote the central values and the shaded bands are the integrated envelopes, see \cref{eqn:xcut_Error}.
We show both results for two choices of the angular dependence: $\beta = 2$ [left] and $\beta = 0.5$ [right].}
\label{fig:C1Lambda}
\end{figure}

In this section, we explore the dependence on the dark sector confinement scale \LambdaD.
The plots shown in \cref{fig:C1Lambda} compare the $\ECF$ distribution for a dark-quark-initiated jet against a QCD-quark-initiated jet for a range of confinement scales $\LambdaD \neq \Lambda_\text{QCD}$ compared to the QCD quark background, for two choices of $\beta$. 
As the confinement scale increases, the dark sector distribution shifts toward larger values of $\ECF$.
The larger confinement scale implies that the dark sector coupling is larger than the QCD coupling at the energy scale of the jet. 
This implies that the peak of the differential distribution occurs at a larger value of $\ECF$, or equivalently, that the resummation approximation $\aD L^2 \sim 1$ becomes relevant for larger values of $\ECF$. 
Therefore, the distribution peaks closer to the kinematic endpoint.

In the bottom row of \cref{fig:C1Lambda}, we provide the cumulative distribution $\Sigma(x_\text{cut})$ for the various choices of \LambdaD.
For $\beta = 2$, the envelope saturates at $x_\text{cut} = 10^{-3}$ for large values of \LambdaD and shifts toward $x_\text{cut} = 10^{-4}$ as \LambdaD decreases.
The range of this envelope is 0.22 and insensitive to the size of \LambdaD.
Similarly, for $\beta = 0.5$, the envelope saturates at $x_\text{cut} = 10^{-2}$.
The envelope range increases as \LambdaD decreases, from a minimum of 0.26 and a maximum of 0.40.

\begin{figure}[t!]
\center
\includegraphics[width=\linewidth]{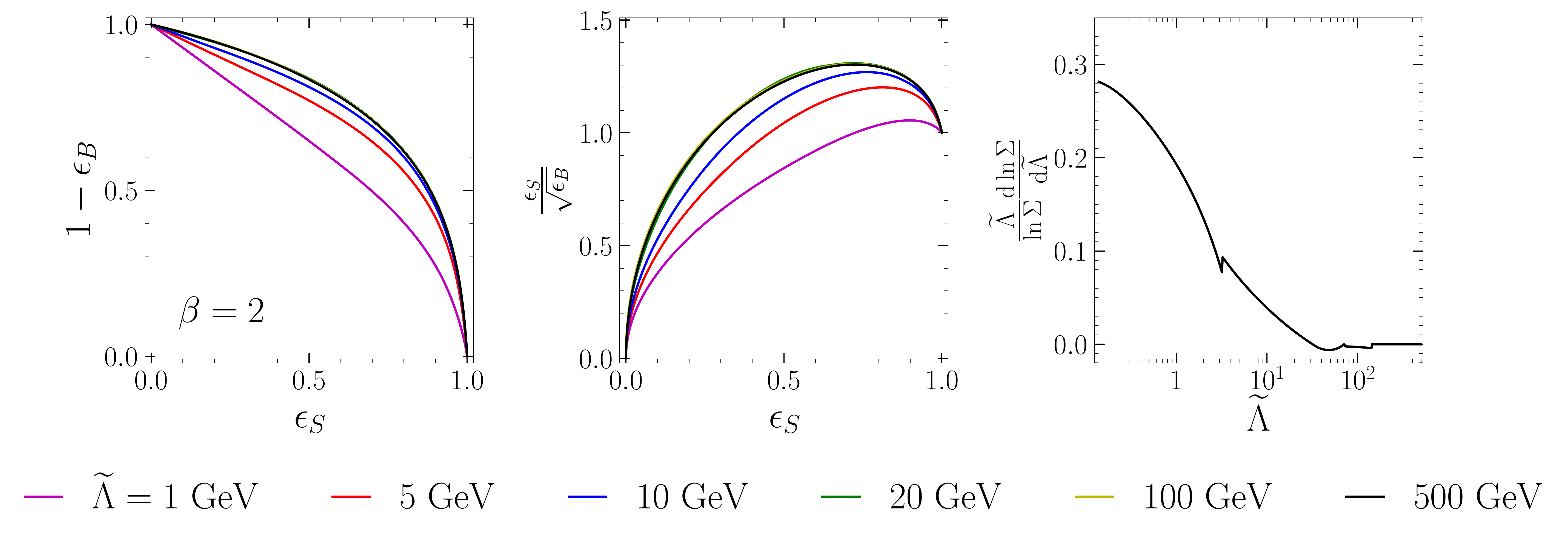}
\caption{Discrimination of dark sector against QCD for various choices of the confinement scale \LambdaD for $p_T = \SI{1}{\TeV}$ jets, using the MLL analytic calculation.
Note that the impact of errors is ignored here, see \cref{sec:Apply} for details.
We show ROC curves in the background rejection $1-\epsilon_B$ versus signal efficiency $\epsilon_S$ plane [left].
We show the curve of discriminatory significance $\epsilon_S/\sqrt{\epsilon_B}$ against signal efficiency $\epsilon_S$ [middle].
Fixing the background rejection at \SI{90}{\percent}, we then show the relative change in discriminatory power as a function of \LambdaD.
The angular dependence parameter is $\beta = 2$ for all panels.
}
\label{fig:C1LambdaDisc}
\end{figure}

As \cref{fig:C1LambdaDisc} shows, the discriminatory power of a dark sector signal against a QCD background increases as the dark sector's confinement scale \LambdaD increases. 
However, this increased discrimination power saturates for large confinement scales $\LambdaD \gtrsim \SI{50}{\GeV}$. 
This saturation is caused by freezing the running coupling at the ``non-perturbative scale'' $\mu_\text{NP} = 7\LambdaD$, which we emphasize is a nonphysical prescription designed to obtain a closed-form solution to \eqref{eqn:radiator}.
Using the explicit dependence of $\mu_\text{NP}$ on $\LambdaD$, we can derive a na\"{i}ve small coupling expansion for the discriminator,
\begin{equation}
\label{eqn:LambdaDisc}
  \frac{\LambdaD}{\ln\Sigma} \dv{\ln\Sigma}{\LambdaD} \simeq
    \left(\ln\frac{p_T R_0}{\LambdaD}\right)^{-1} \,.
\end{equation}
This provides a reasonable estimate of the scaling until $\LambdaD \gtrsim \SI{5}{\GeV}$, when the approximation begins to fail.
This can be traced back to the behavior of \cref{eqn:radiator_g1_A,eqn:radiator_l1_A,eqn:radiator_e1_A}, from which we infer that as the confinement scale increases, the non-perturbative effects become more relevant for larger values of $\ECF$.

\begin{figure}[t!]
\center
\includegraphics[width=\linewidth]{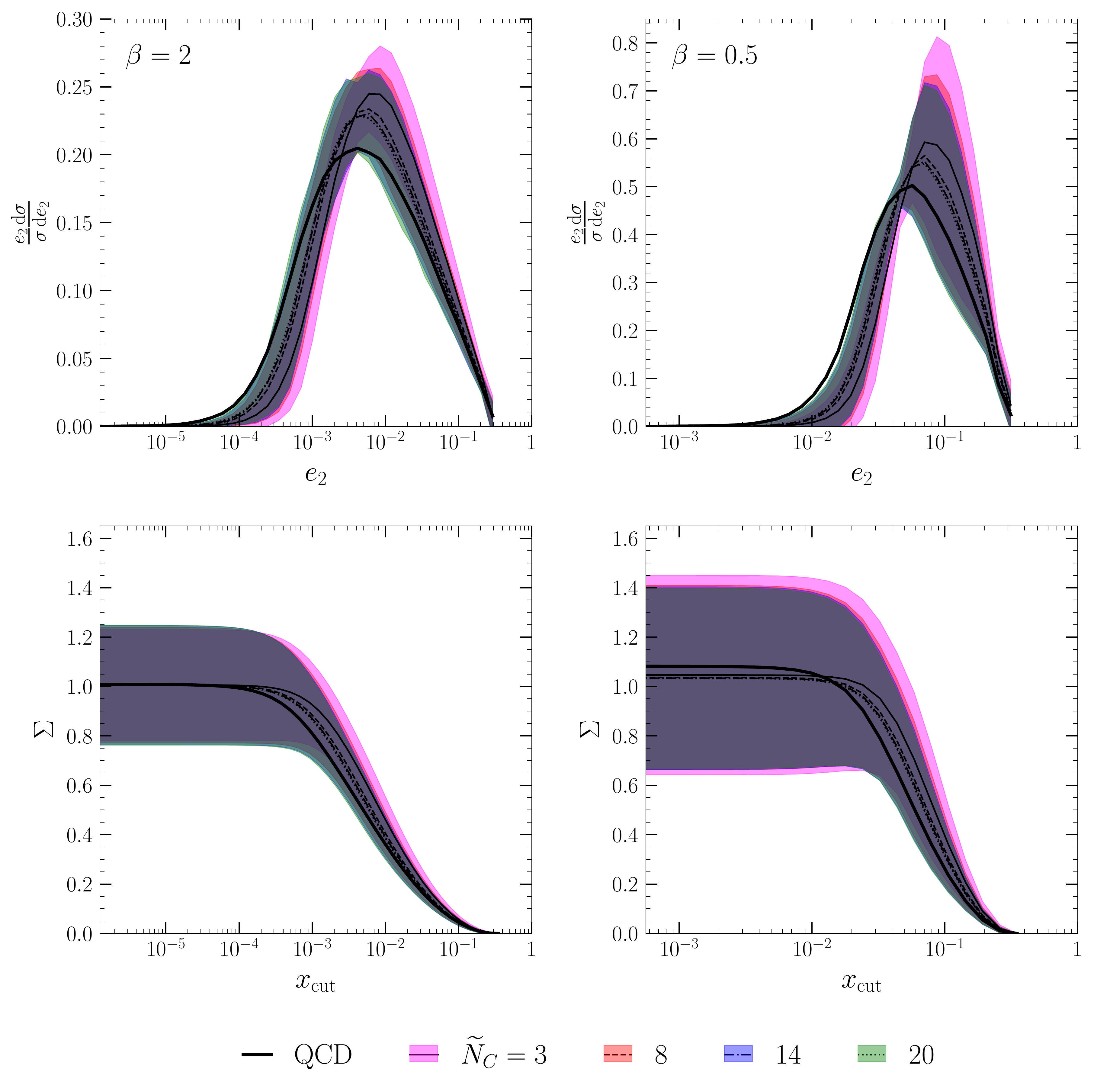}
\caption{The behavior of \ECF as the dark sector gauge group $SU(\ncD)$ is varied, for $p_T = \SI{1}{\TeV}$ jets.
See the legend for values of \ncD; all other values are given in \cref{tab:hiddenparams}.
We show the normalized \ECF distributions [top], where the central value of the envelope is marked with the black lines, while the shaded region denotes the envelope.
The peak moves to slightly lower values of \ECF as \ncD is increased.
The cumulant distributions $\Sigma$ as a function of $x_\text{cut}$ are also provided [bottom], where again the lines denote the central values and the shaded bands are the integrated envelopes, see \cref{eqn:xcut_Error}.
We show both results for two choices of the angular dependence: $\beta = 2$ [left] and $\beta = 0.5$ [right].}
\label{fig:C1Color}
\end{figure}

%**************** Subsection *******************************
\subsection{\boldmath \ncD Dependence}
%*************************************************************
In this section, we explore how the substructure depends on the number of dark colors, \ncD.
The set of plots shown in \cref{fig:C1Color} compare the $\ECF$ distribution for a QCD-quark-initiated jet against a dark sector-quark-initiated jet for various choices of the number of dark colors $\ncD > 3$. 
As the number of dark colors increases, the $\beta$-function for the dark sector gauge coupling becomes more negative, so the scale evolution is faster for the dark sector than for the QCD background. 
This faster running of the coupling shifts the dark sector distribution toward smaller values of $\ECF$ since $\aD < \aS$ at the scale set by the jet $p_T$.

In the bottom row of \cref{fig:C1Color}, we provide the cumulative distribution $\Sigma(x_\text{cut})$ for the various choices of \ncD.
For $\beta = 2$, the envelope saturates at $x_\text{cut} = 10^{-4}$, regardless of the value of \ncD.
The range of this envelope is 0.22 and insensitive to the size of \ncD.
Similarly, for $\beta = 0.5$, the envelope saturates at $x_\text{cut} = 10^{-2}$.
The envelope range increases as \ncD decreases, from a minimum of 0.36 and a maximum of 0.40.

\begin{figure}[t!]
\center
\includegraphics[width=\linewidth]{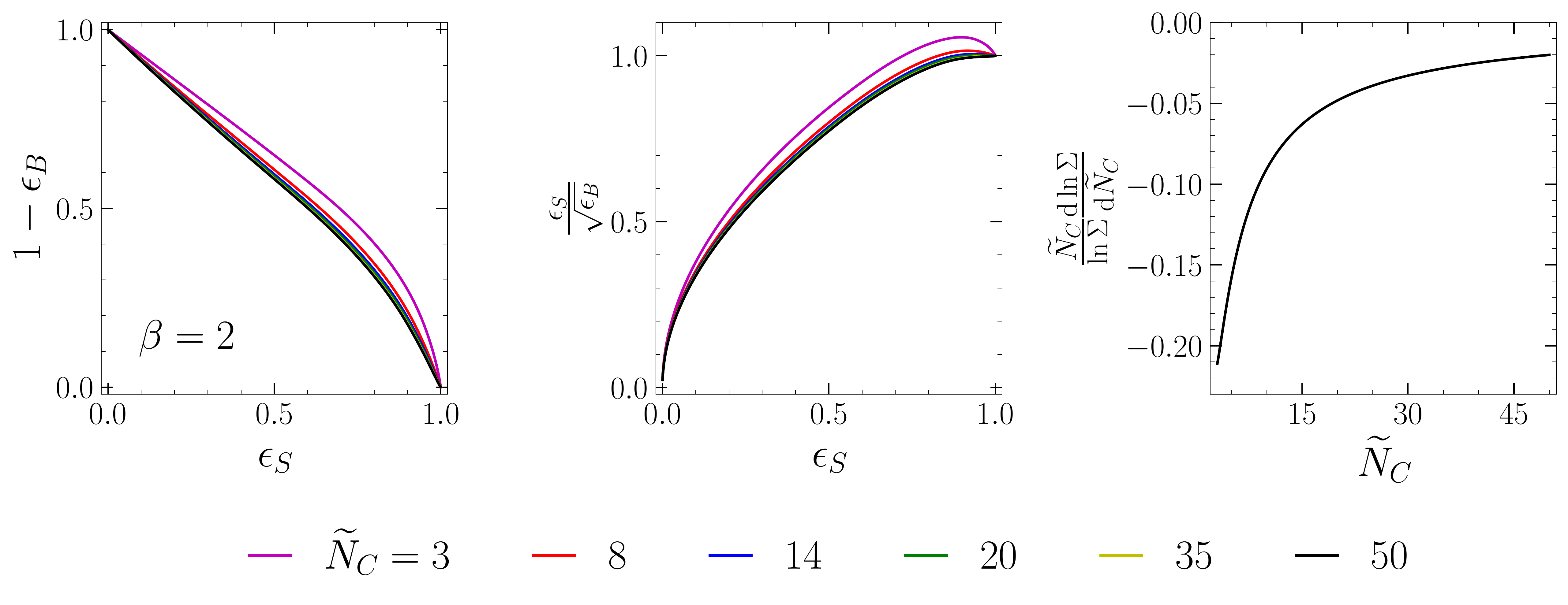}
\caption{Discrimination of dark sector against QCD for various choices of the number of colors in the dark sector \ncD for $p_T = \SI{1}{\TeV}$ jets, using the MLL analytic calculation.
Note that the impact of errors is ignored here, see \cref{sec:Apply} for details.
We show ROC curves in the background rejection $1-\epsilon_B$ versus signal efficiency $\epsilon_S$ plane [left].
We show the curve of discriminatory significance $\epsilon_S/\sqrt{\epsilon_B}$ against signal efficiency $\epsilon_S$ [middle].
Fixing the background rejection at \SI{90}{\percent}, we then show the relative change in discriminatory power as a function of \ncD.
The angular dependence parameter is $\beta = 2$ for all panels.
}
\label{fig:C1ColorDisc}
\end{figure}

As \cref{fig:C1ColorDisc} shows, the discriminatory power of a dark sector signal against the QCD background decreases as the dark sector's number of dark colors \ncD increases. 
However, this decrease is rather marginal, and saturates for $\ncD \sim 10$. 
We can understand this behavior analytically, by expanding the LL resummed cumulative distribution \cref{eqn:radiatorLL_A} to leading order in \aD.
We find that the \ncD dependence is well approximated by
\begin{equation}
\label{eqn:ColorDisc}
  \frac{\ncD}{\ln\Sigma} \dv{\ln\Sigma}{\ncD} \simeq
    \frac{1+\ncD^2}{\ncD^2-1} + \frac{11\ncD}{4\nfD T_R-11\ncD} \,.
\end{equation}
This makes it clear that the discriminator quickly asymptotes as one increases \ncD, thereby explaining the qualitative behavior in the figures, \ie, that the sensitivity of the observables studied here to the number of dark colors is minimal.
%\newpage

%**************** Subsection *******************************
\subsection{\boldmath \nfD Dependence}
\begin{figure}[t!]
\center
\includegraphics[width=\linewidth]{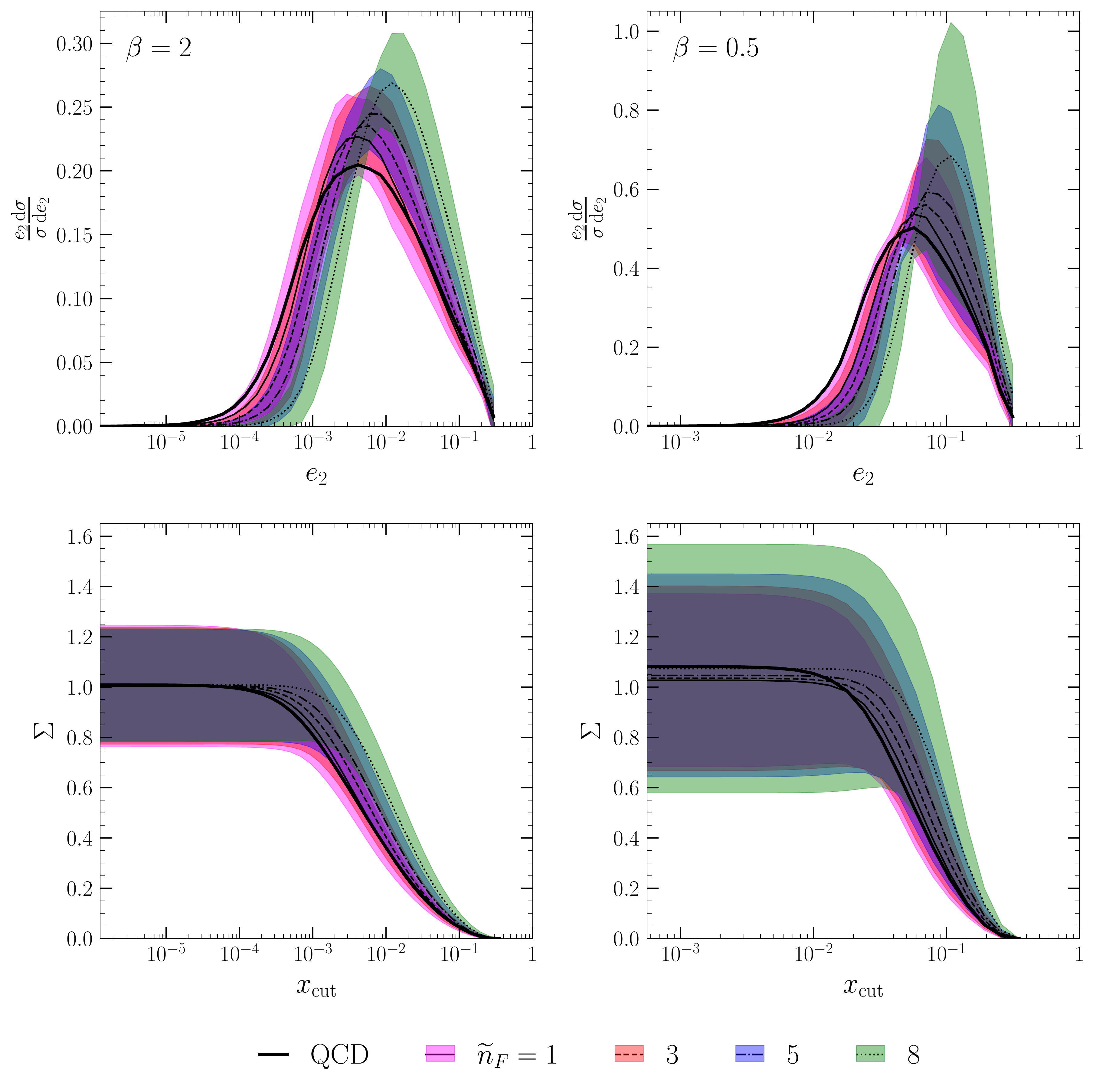}
\caption{The behavior of \ECF as the number of dark quark flavors \nfD is varied, for $p_T = \SI{1}{\TeV}$ jets.
See the legend for values of \nfD; all other values are given in \cref{tab:hiddenparams}.
We show the normalized \ECF distributions [top], where the central value of the envelope is marked with the black lines, while the shaded region denotes the envelope.
The peak moves to higher values of \ECF as \ncD is increased.
The cumulant distributions $\Sigma$ as a function of $x_\text{cut}$ are also provided [bottom], where again the lines denote the central values and the shaded bands are the integrated envelopes, see \cref{eqn:xcut_Error}.
We show both results for two choices of the angular dependence: $\beta = 2$ [left] and $\beta = 0.5$ [right].}
\label{fig:C1Flavor}
\end{figure}

%*************************************************************
In this section, we explore how the substructure depends on the number of dark flavors \nfD.
The plots shown in \cref{fig:C1Flavor} compare the $\ECF$ distribution for a dark-quark-initiated jet against a QCD-quark-initiated jet for a range of dark flavors with $\nfD > 5$; note that we take the number of flavors for QCD to be $n_F = 5$. 
As the number of dark flavors increases, the $\beta$-function for the dark sector coupling \aD decreases, and in particular the dark sector is no longer asymptotically free when $\nfD > \frac{11\ncD}{4T_R}$.
This implies that the renormalization group evolution is slower for the dark sector than for the QCD background. 
This impacts the dark sector distribution by shifting it towards larger values of $\ECF$, since $\aD > \aS$ at the characteristic hard scale of the jet.

In the bottom row of \cref{fig:C1Flavor}, we provide the cumulative distribution $\Sigma(x_\text{cut})$ for the various choices of \nfD.
For $\beta = 2$, the envelope saturates at $x_\text{cut} = \num{5e-4}$ for large values of \nfD and shifts toward $x_\text{cut} = 10^{-4}$ as \nfD decreases.
The range of the envelope is 0.24 and insensitive to the size of \nfD.
Similarly, for $\beta = 0.5$, the envelope saturates at $x_\text{cut} = 10^{-2}$, regardless of the value of \nfD.
The envelope range increases as \nfD increases, from a minimum of 0.34 and a maximum of 0.50.
While we are limited by how many flavors we allow the dark sector to have if we want a confining dark sector, the differential distribution shifts toward larger values of \ECF as \nfD increases.

\begin{figure}[t!]
\center
\includegraphics[width=\linewidth]{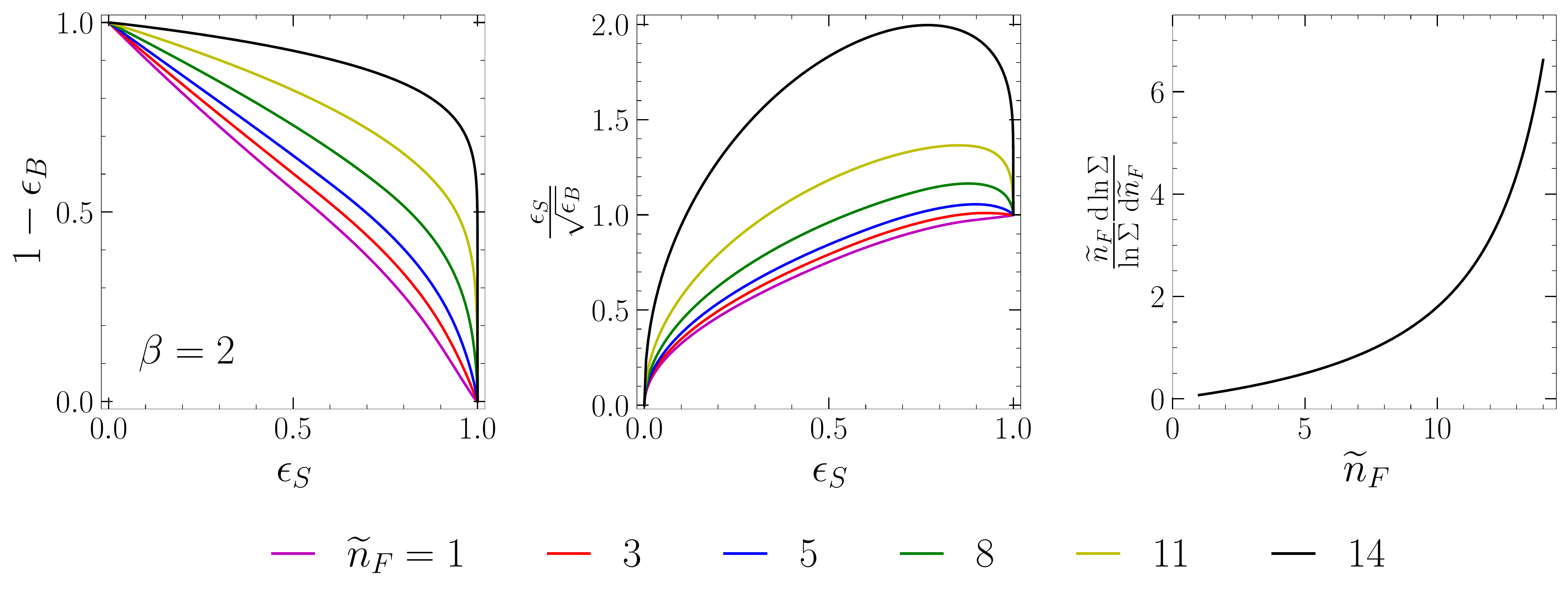}
\caption{Discrimination of dark sector against QCD for various choices of the number of dark quark flavors \nfD for $p_T = \SI{1}{\TeV}$ jets, using the MLL analytic calculation.
Note that the impact of errors is ignored here, see \cref{sec:Apply} for details.
We show ROC curves in the background rejection $1-\epsilon_B$ versus signal efficiency $\epsilon_S$ plane [left].
We show the curve of discriminatory significance $\epsilon_S/\sqrt{\epsilon_B}$ against signal efficiency $\epsilon_S$ [middle].
Fixing the background rejection at \SI{90}{\percent}, we then show the relative change in discriminatory power as a function of \nfD.
The angular dependence parameter is $\beta = 2$ for all panels.
}
\label{fig:C1FlavorDisc}
\end{figure}

As \cref{fig:C1FlavorDisc} shows, the ability to discriminate a dark sector signal against a QCD background increases as the number of dark flavors increases. 
Furthermore, this effect increases rapidly as $\nfD^{-1}$. 
The dark flavor dependence can be estimated by expanding the LL resummed cumulative distribution given in \cref{eqn:radiatorLL_A} to leading order in the coupling.
This yields
\begin{equation}
\label{eqn:FlavorDisc}
  \frac{\nfD}{\ln\Sigma} \dv{\ln\Sigma}{\nfD} \simeq
    \frac{4\nfD T_R}{11\ncD - 4\nfD T_R}\,.
\end{equation}
While naively this implies that we should be able to find regions of parameter space that are very non-QCD-like, the framework breaks down for $\nfD > \frac{11\ncD}{4T_R}$, because the dark sector does not confine as mentioned above.
Practically, \Pythia has limited the number of dark flavors one can include to be eight at most.
Therefore, we are not able to numerically probe the discriminator beyond this point in parameter space.
However, the trend agrees between the numeric and analytic calculations, and follows the analytic estimate in \cref{eqn:FlavorDisc} to a good approximation.

%**************** Subsection *******************************
\subsection{\boldmath $\mD$ Dependence}
%*************************************************************
Finally, we explore the impact of varying \mD on the \ECF distribution.  
Since the analytic calculations assume massless partons, we are not in a position to include the analytic contributions to our error envelopes.
However, for IRC-safe observables such as \ECFb, the mass dependence of our distributions is suppressed as a power of $\mD/\LambdaD$ when $\mD \ll \LambdaD$, with the resulting effect on our results being negligible.
This is not the case when quark masses exceed the confinement scale since $\mD$ then set the scale where the parton shower terminates.
In the latter case, an accurate analytic treatment of finite quark masses is challenging, due to the presence of multiple overlapping logarithms of both $\ECF$ and ratios of quark masses and energy scales.
As a result, the resummation of differential distributions becomes a more involved procedure, and we will content ourselves with simply providing the results of a numerical study, and will not estimate the error band for different choices of \mD.

With a degenerate spectrum, the impact of finite dark quark masses within \Pythia is limited at stopping the parton shower from emitting at scales below \mD, since the resulting partons would not be able to subsequently hadronize, and treating the color strings as having massive endpoints in the evolution of the Lund string during the hadronization step.
Since gluon splitting to quark pairs is not included, potential finite mass effects due to radiation dead cones around additional massive quarks from $g \rightarrow q \bar{q}$ splitting play no role.
Matrix element corrections in emission, which induce additional mass-dependence in analogous QCD showers, are not included.

\begin{figure}[t!]
\center
\includegraphics[width=\linewidth]{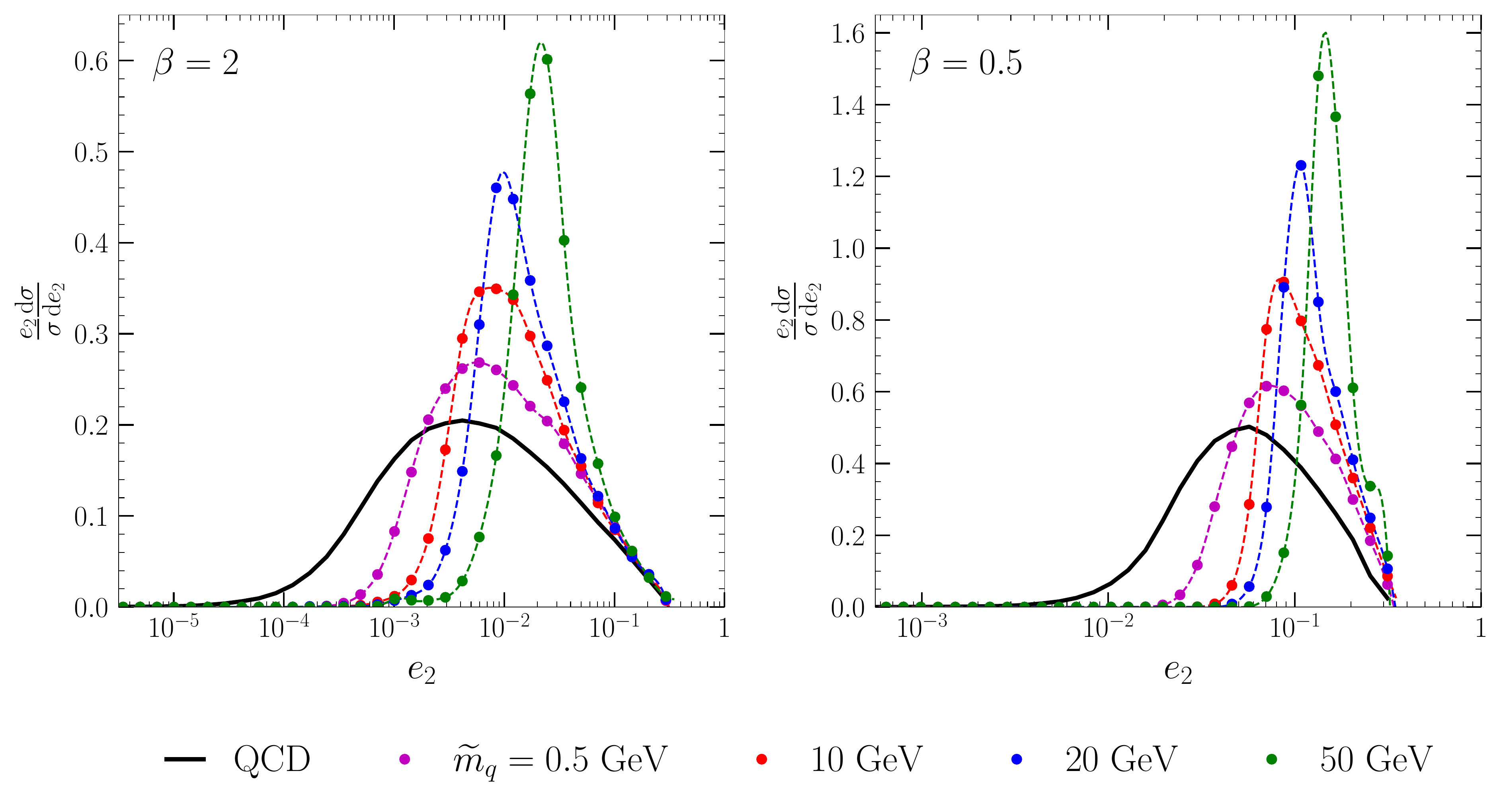}
\caption{The behavior of \ECF for the dark sector as the degenerate dark quark mass \mD is varied, for $p_T = \SI{1}{\TeV}$ jets.
The value of \mD is varied according to the legend, while all other values are given in \cref{tab:hiddenparams}.
The associated dark hadron masses are 2\mD.
Only a numerical study using \Pythia is presented.
We provide a cubic fit to these distributions to guide the eye.
Larger dark quark masses move the peak to higher values due to the resulting cutoff imposed on collinear divergences for emissions from massive quarks.
We show the results for two choices of the angular dependence: $\beta = 2$ [left] and $\beta = 0.5$ [right].
}
\label{fig:C1Mass}
\end{figure}

When the quark masses are above the confinement scale, the hadrons are more akin to quarkonia like the $J/\psi$ or $\Upsilon$ than they are to light mesons like the $\pi$ or $\rho$.
Hadronization still occurs, since individual dark quarks cannot decay and can only annihilate once they become bound into hadrons.
While the properties of these states may be well approximated by perturbative methods, as long as the multiplicity of quarkonia produced is larger than a few, a parton shower is still expected to provide a good approximation of the final state.\footnote{This approximation will break down as the mass of the dark quarks approaches the energy of the jet such that production of multiple hadrons becomes kinematically disfavored. In this limit, dark glueball production would be expected to result in additional energy missing from the final state. We make no attempt to model this effect.}

The result is displayed in \cref{fig:C1Mass}, where we compare the \ECF distribution for a quark-initiated QCD jet against the \Pythia distributions for different choices of the dark quark mass \mD; we assume that the dark quarks are degenerate and that the dark hadron masses are 2\mD for simplicity.
The other dark sector parameters are set to the default given in \cref{tab:hiddenparams}.
We see that the peak of the distributions moves to higher values of \ECF as \mD is increased.
Additionally, we note that the impact on the distributions is not as dramatic as when we varied \LambdaD above (\emph{cf}., \cref{fig:C1Lambda}).
This can be understood due to the fact that increasing the quark masses for fixed gauge coupling simply acts to cut out more of the IR region of the shower phase space where the sector is becoming strongly coupled.
While this has an impact on the resultant multiplicity of dark hadrons that are produced in a shower, their subsequent decay from a higher rest mass to nearly massless QCD hadrons obscured the impact of the specific mass scale set by \mD on the observable distribution.

%**************** Section *******************************
\section{Quantifying Hadronization Uncertainties}
\label{sec:Had}
%*************************************************************
The enveloping procedure includes variations among predictions that result from either an analytic or a numerical approach to capture the dominant IR logs that result from showering.
When considering sources of systematic uncertainties, it is critical to investigate the irreducible error on predictions due to incalculable strong coupling effects.
Specifically, the numerical results rely on a phenomenological model of hadronization.
In the case of \Pythia, the hadronization step uses the Lund string model~\cite{Andersson:1983ia}, which models the physics of confinement by iteratively connecting partons to each other with color strings, and breaking these strings by pair producing quarks from the vacuum when energetically favorable until an equilibrium configuration is achieved.\footnote{Another commonly used Monte Carlo shower program is \Herwig~\cite{Bellm:2015jjp}, which uses the cluster hadronization model.}
This approach introduces incalculable parameters, which can be tuned to data in the case of real QCD, but must simply be set by hand for the dark sector.
It is therefore critical to our goals here to include the uncertainty associated with these choices.
As we will show here, hadronization systematics are of the same size as the perturbative ones included in the error envelopes thus far.
Clearly, they should additionally be included for searches performed by the experimental collaborations.

\label{sec:varyHad}
\begin{figure}[t!]
\center
\includegraphics[width=\linewidth]{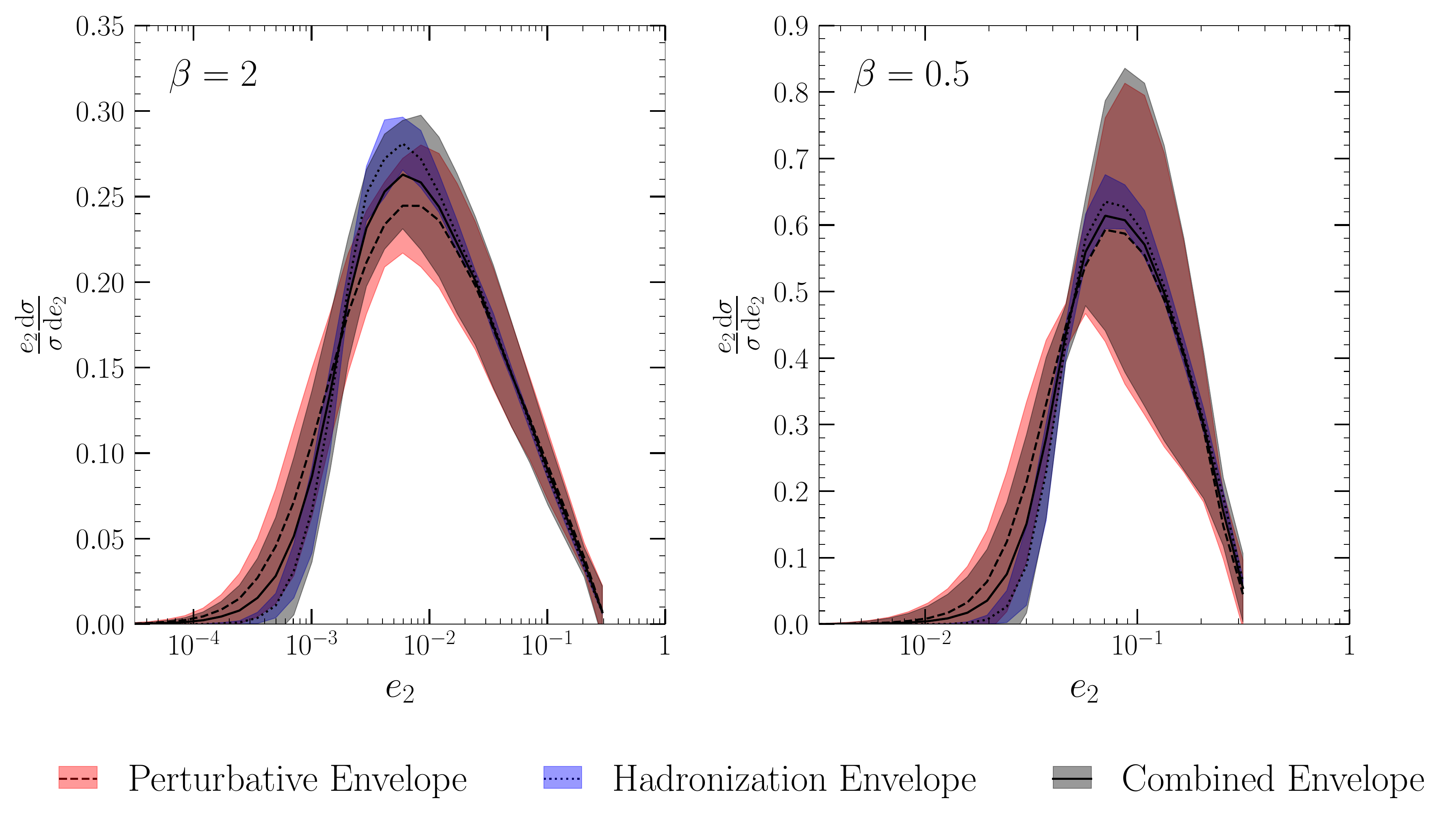}
\caption{Total theory uncertainties on the \ECF distributions for $p_T = \SI{1}{\TeV}$ jets for a dark sector whose model parameters are set to the benchmark values given in~\cref{tab:hiddenparams} due to both perturbative and hadronization effects.
Perturbative uncertainties are largest around the peak region, dominating at larger \ECF values.
Hadronization uncertainties contribute most noticeably starting from the peak and extend down to smaller values of \ECF.
Results are presented for two choices of the angular dependence: $\beta = 2$ [left] and $\beta = 0.5$ [right].
}
\label{fig:C1Lund}
\end{figure}

The results of varying the hadronization parameters is given in \cref{fig:C1Lund}, where all other dark sector model parameters are set to the benchmark values given in~\cref{tab:hiddenparams}.
We then explored the hadronization parameter space to find a choice that resulted in the least (most) number of dark hadrons, which corresponds to the parameter choices $\texttt{aLund} = 0$, $\texttt{bmqv2} = 2$, and $\texttt{rFactqv} = 0$ ($\texttt{aLund} = 2$, $\texttt{bmqv2} = 0.2$, and $\texttt{rFactqv} = 2$).
The hadronization band in \cref{fig:C1Lund} is then computed by taking the envelope across the result of the default hadronization parameters and these two extreme choices.
For reference, we also plot the perturbative prediction, and provide the error envelope as computed above with default hadronization parameters, and we also show the combination of the two envelopes by adding them in quadrature.
The largest impact is that the peak of these distributions do shift; this is expected since the position of the turn over is not under robust theoretical control. 
We see that the variation from hadronization is of the same order as the perturbative uncertainty.\footnote{We caution that other observables could be even more sensitive to the details of hadronization, especially those that rely on shape aspects of the substructure.}
We will use the total envelope in the next section where we estimate the impact of non-trivial error envelopes on a mock search for dark sector substructure.

%**************** Section ***********************************
\section{Discovering Dark Substructure}
\label{sec:discovery}
%*************************************************************
Having quantified the perturbative and hadronization theory errors on the prediction for substructure that results from dark sector showering, we will briefly turn to estimating the impact of including our error envelopes for a search.
Our goal here is to simply estimate the discovery potential.
Unsurprisingly given existing limits, the subtle nature of the signature and the overwhelming size of the QCD background will imply that additional handles are required to reduce the background by a factor of $\ord(10^5)$ if there is any hope of seeing evidence for dark substructure signals.
For example, in models where some of the dark hadrons are stable, a cut on missing energy could play this role.
In this case, ignoring the effect of jet to jet fluctuations in the number of unstable mesons, the predictions made above are unchanged, except that the statistics are reduced due to the fact that some particles are missing.
We expect the associated theory uncertainty to be a subleading effect.

One important mitigating factor is that stringent limits on new physics contributions to QCD distributions already exist from ATLAS~\cite{Aaboud:2017yvp} and CMS~\cite{Sirunyan:2018wcm}.
Since these searches simply look for high $p_T$ jets in the final state, the dark jets would fall in the signal region with essentially equal efficiency to QCD jets.
Therefore, our first step to quantifying the discovery reach for models that yield substructure from dark showers is to interpret these bounds as a limit on the dark quark production cross section.

We assume the portal to the dark sector can be modeled by a contact interaction:
\begin{equation}
\label{eqn:contact}
 \mathcal{L}_\text{int} \supset
   \frac{1}{\Lambda_{\text{CI}}^2} (\bar{q}\s\gamma^\mu\s q) (\bar{\widetilde{q}}\s\gamma_\mu\s \widetilde{q}) \,.
\end{equation}
By hunting for deviations in the tails of jet distributions, ATLAS~\cite{Aaboud:2017yvp} and CMS~\cite{Sirunyan:2018wcm} have derived comparable limits $\Lambda_{\text{CI}} \gtrsim \SI{22}{\TeV}$. 
We emphasize that this limit is essentially unchanged for our model, since the searches do not make any cuts on substructure.

We convert this limit on the new physics scale into a bound on the production cross section using an implementation of a $B-L$ extension of the Standard Model~\cite{Basso:2008iv, Deppisch:2018eth} publicly available in the \textsc{FeynRules}~\cite{Alloul:2013bka} model database.
We take the $Z'$ mass to be large so that the production process $\bar{q}\s q \to Z' \to \bar{\widetilde{q}}\,  \widetilde{q}$ is well approximated by \cref{eqn:contact}.
Events are simulated using \textsc{MadGraph5\_aMC@NLO}~\cite{Alwall:2014hca}, taking the model parameters to correspond to the lower bound on $\Lambda_{\text{CI}}$.
This allows us to compute the cross section for $p\s p\to \overline{q}\s q$, and we then simply interpret the result as the rate for dark quark production.
We implement generator level cuts on rapidity $\eta < 2$ and transverse jet momentum $p_{T} > \SI{1}{\TeV}$.
Our dijet backround is produced by all $2 \to 2$ QCD processes applying the same cuts.
This results in a signal cross section $\sigma_S = \SI{5E-5}{\pb}$, which can be compared to the enormous QCD background $\sigma_B = \SI{13}{\pb}$.\footnote{Since this is meant to be a simple estimate, we do not include a $K$-factor, which at NNLO is in the range $1.3-1.5$~\cite{Currie:2017eqf}.}
These cross sections are used to compute the expected number of events for two choices of integrated luminosity; the final Run~III data set of \SI{300}{\per\fb} and the complete high luminosity data set of \SI{3000}{\per\fb}. 
These values should be interpreted as the number of events that survive a loose ``pre-selection'' for the search.

Next we approximate the discovery significance including the impact of both statistical and systematic uncertainties using
\begin{equation}
\label{eqn:significance}
  \mathcal{Z} = \frac{S}{\sqrt{S+B+\delta_S^2 S^2 + \delta_B^2 B^2}}\,,
\end{equation}
where $S$ is the number of signal events, $B$ is the number of background events, and $\delta_i$ are their respective systematic uncertainties. 
Given the already stringent limits on the production of the dark quarks, it is easy to check that using dark substructure alone will not provide enough discriminating power to beat down the QCD background.
Therefore, we will reframe the question in terms of a background reduction factor $\epsilon$, which provides an estimate of what one must be able to achieve by incorporating other handles into the search, \eg~missing energy, resonances, and/or displaced objects.\footnote{These changes to the model would obviously also impact the limits on signal production rates, \ie, the limit on $\Lambda_{\text{CI}}$ in \cref{eqn:contact}.}
To compute $\epsilon$, we solve \cref{eqn:significance} using the substitution $B \to \epsilon\s B$:
\begin{equation}
\label{eqn:reduction}
  \epsilon = \frac{\sqrt{1-4 \delta_B^2 S\bigl(1+\bigl(\delta_S^2-\frac{1}{\mathcal{Z}^2}\bigr)S\bigr)}-1}{2\delta_B^2\s B}\,.
\end{equation}
Larger values of $\epsilon$ correspond to improved discrimination.

\begin{figure}[t!]
\center
\includegraphics[width=\linewidth]{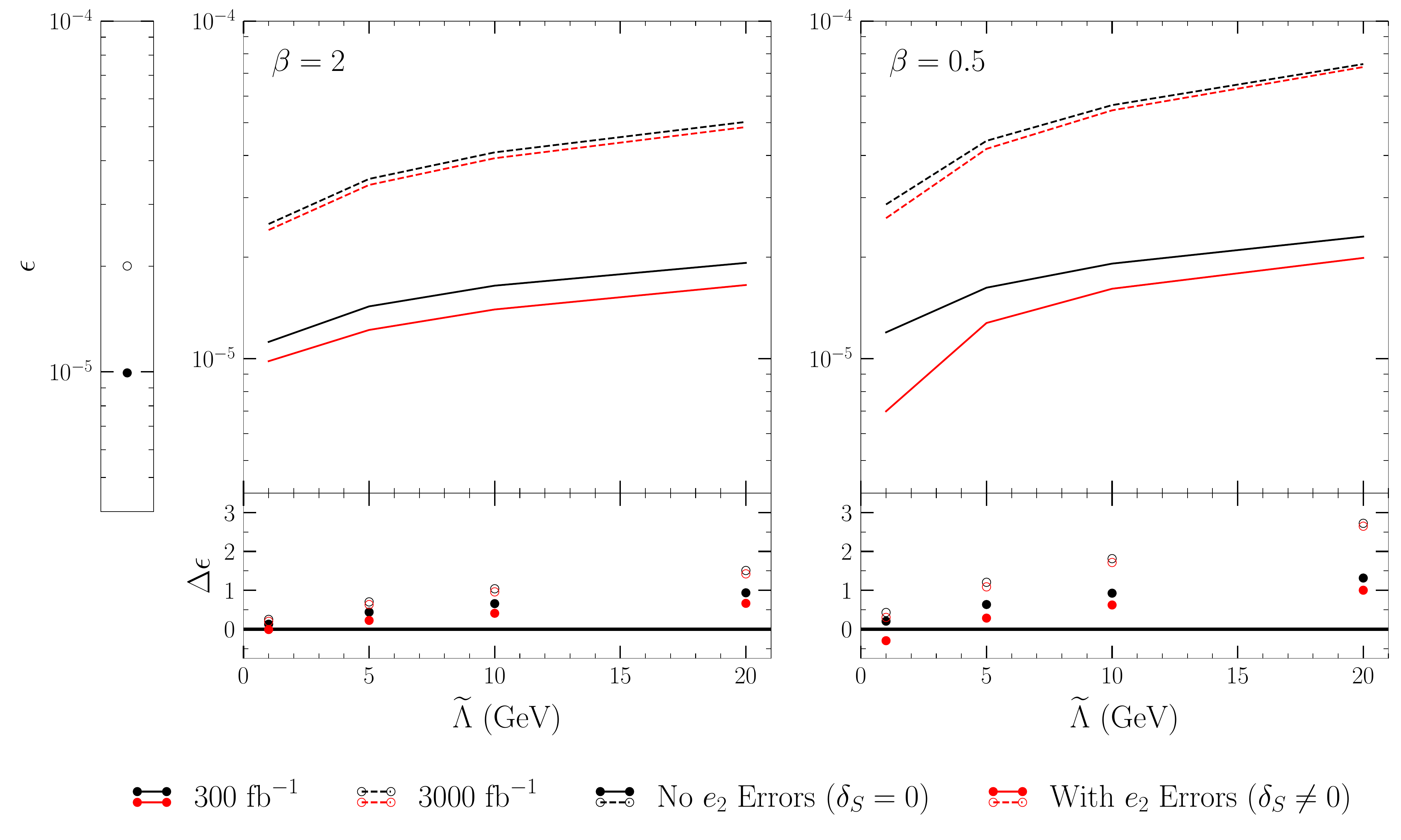}
\caption{The background reduction factor (see \cref{eqn:reduction}) required to observe new physics over the QCD background for $p_T = \SI{1}{\TeV}$ jets.
We show the value required if there are no additional cuts made on jet substructure (assuming $\delta_S = 0$) [left].
Then we provide the results as a function of \LambdaD taking $\beta = 2$ [middle] and $\beta = 0.5$ [right].
The resulting reduction in the needed $\epsilon$ as a result of the substructure cut is presented as a multiplicative factor below the middle and right plots.
We vary the luminosity and provide results with and without errors, see the legend for details.
}
\label{fig:C1LambdaSig}
\end{figure}

First, we estimate how large $\epsilon$ would need to be in order to see a $2\sigma$ excess of signal events \emph{without} a cut on substructure and assuming no uncertainty on the signal production rate and assuming the cut has no impact on signal statistics, see the left panel of \cref{fig:C1LambdaSig}.
This provides a baseline against which we can compare how much improvement can be obtained using substructure.
Next, we include the substructure cut, using the models with varying $\LambdaD$ as a concrete example.
We assume the theory error bands on the dark sector distributions are fully correlated, just as we did above when computing the cumulative distributions, \eg~\cref{fig:C1Lambda}.
For the QCD background, there is a wealth of data that is used for tuning and calibration, and as such the systematic error bars can be controlled by leveraging a variety of inputs.
For the results presented in \cref{fig:C1LambdaSig}, we use the background uncertainty $\delta_B = 0.1$ as determined by a recent NNLO calculation~\cite{Currie:2017eqf}.
We additionally assume that $\delta_B$ does not depend on the substructure cut.
As a point of comparison, data driven approaches currently yield $\delta_B \sim 20\%$~\cite{Aaboud:2017yvp}.

In \cref{fig:C1LambdaSig}, we plot the background rejection factor required to achieve a $2\sigma$ exclusion as a function of the dark confinement scale \LambdaD, by optimizing a substructure cut for each choice of model parameters.
In order to explore the impact of the error envelopes, we provide the result with $\delta_S=0$ in black and $\delta_S \neq 0$ in red, and we also provide the results for $\beta = 2$ and $0.5$ to investigate varying the angular dependence parameter.
We assume either \SI{300}{\per\fb} or \SI{3000}{\per\fb} of integrated luminosity, which allows us to explore the scaling as the data set size is increased.\footnote{It is worth noting that the bound on the scale for the contact operator $\Lambda_{\text{CI}}$ will also improve with more data, which is not being taken into account here.}

Most importantly, we see that a cut on substructure improves one's ability to discover these models, even when the systematic error on the signal shape is included.
In particular, taking $\beta = 2$ and $\LambdaD = 20 \text{ GeV}$ the relative change $\Delta\epsilon = 0.9 (0.6)$ for no error (with error) for \SI{300}{\per\fb}; the relative change for \SI{3000}{\per\fb} $\Delta\epsilon = 1.5 (1.4)$ for no error (with error).\footnote{Note that we have not included the constraint on the error of the integrated resummed cross section given by the inclusive calculation. This is the cause of the unphysical value for the background reduction factor for $\LambdaD = \SI{1}{\GeV}$ and $\beta = 0.5$. This only becomes a significant effect when the additional sensitivity gained from substructure discriminants becomes small.}
This motivates future work quantifying the error envelopes for a wider variety of substructure distributions that could result from dark sector showers, so that cuts on these variables can be properly incorporated into searches.
In particular, it is important to include such systematics when deriving limits on signal parameter space, since the non-trivial error bands can result in more realistic exclusion regions.
Finally, we note that for the \SI{300}{\per\fb} data set, the optimized value of the cut yields a signal region that is statistics dominated.
Then when we increase the data set size to \SI{3000}{\per\fb}, we find that optimal signal region has comparable statistical and systematic errors.
We conclude that this subtle signature of dark sector physics is an interesting target scenario for the physics program at the high luminosity LHC.

%**************** Section ***********************************
\section{Conclusions}
\label{sec:end}
%*************************************************************
In this paper, we explored the theory uncertainties associated with making predictions for a scenario where the presence of a new strongly coupled dark sector leaves its imprint on the substructure of QCD-like jets.
We focused on the two-point energy correlation function, \ECFb.
In particular, we quantified the error resulting from perturbative uncertainties associated with truncating to finite order in the logarithmic and gauge coupling expansions.
We also explored the uncertainty due to incalculable non-perturbative hadronization effects.
Varying the dark confinement scale \LambdaD had the most pronounced impact on the shape of the resulting distributions.
We showed \ECFb to be relatively insensitive to the number of dark colors \ncD but observed more striking variations when varying the number of dark flavors \nfD.
We also briefly explored the dependence on the dark quark mass, although we did not provide an error envelope for these distributions due to the technical limitations discussed above.

We then used these error estimates to quantify one's ability to distinguish dark sector jets from the QCD background.
We assumed that current bounds on four-quark contact operators apply, which was used to set the production rate for the dark sector.
Achieving sensitivity to this subtle signal requires introducing additional handles for the search strategy that could reduce the QCD background by a factor of $\ord(10^5)$ assuming little impact on the signal efficiency.
Depending on the model, one could implement a cut on missing energy, a requirement of one or more $b$-tagged jets, or identification of displaced vertices or resonances --- these additional uncorrelated features could additionally impact the interpretation of the limit on the production cross section, a full exploration of the open parameter space for variations of the base model is an interesting topic for future work.\footnote{If these features are in fact correlated, then the analysis performed in this paper would need to be redone.}
This signature may also provide an interesting target opportunity for model agnostic approaches to new physics searches that rely on machine learning, \eg~\cite{Farina:2018fyg, Heimel:2018mkt, Collins:2019jip, DAgnolo:2019vbw, Bradshaw:2019ipy, Nachman:2020lpy, Andreassen:2020nkr, Hajer:2018kqm}.
While such approaches could mitigate the impact of theory uncertainties on the discovery potential of searches using substructure, the importance of uncertainties in setting accurate limits or extracting model parameters in the case of discovery cannot be ignored.
Regardless of these details, this study makes clear that a dedicated search that relies on subtle features in substructure will benefit from the full data set collected at the high luminosity LHC, thereby providing a compelling physics target for future experimental efforts.

Moving forward, we acknowledge the practical need for the generalization of the error envelopes presented here to additional substructure variables.
It is important to note that properly accounting for the impact of theory errors for a different observable of interest would require a similar study to what we have presented above.
In particular, comparable analytic calculations are necessary to characterize theory uncertainties.
We do expect that for a class of mass-like observables, \ie, those that display Casimir scaling at LL~\cite{Larkoski:2013eya}, one would find conclusions broadly similar to the case of \ECF.
However, there are cases, \eg~those briefly mentioned at the beginning of \cref{sec:frame}, with a sufficiently different structure, such that a dedicated study would be necessary to determine the size and scaling of the errors.
In the case of uncertainties which can be reliably characterized via Monte Carlo alone, \eg~hadronization modeling, modern machine learning methods similar to those of \Ref{Andreassen:2019nnm} might prove helpful in reducing the effort involved.
However, we emphasize that a proper analytic accounting of expected theory errors in a resummed calculation has no true substitute.
The work presented here makes the case that a comprehensive characterization of how substructure observables can be most useful for LHC applications should be performed.

\acknowledgments
We thank Deepak Kar, Graham Kribs, Simone Marzani, Bryan Ostdiek, Kevin Pedro, Sukanya Sinha, and Dave Soper for useful discussions. We are particularly grateful to Simone Marzani for sharing his numerical implementation of the analytic expressions from \Ref{Larkoski:2014wba}.  TC and JD are supported by the U.S. Department of Energy (DOE) under grant DE-SC0011640.  MF is supported by the DOE under grant DE-SC0010008 and partially by the Zuckerman STEM Leadership Program. TC and MF performed some of this work at the Munich Institute for Astro- and Particle Physics (MIAPP) which is funded by the Deutsche Forschungsgemeinschaft (DFG, German Research Foundation) under Germany's Excellence Strategy -- EXC-2094 -- 390783311. MF thanks the Berkeley Center for Theoretical Physics and Lawrence Berkeley National Laboratory for their hospitality during the completion of this work.

\appendix

%*********************************** Section ***********************************%
\section{Analytic Calculation}
\label{sec:analytics}
%*********************************** *******************************************%
%
In this appendix, we review the analytic calculation of the \ECF distribution to NLL order. 
Our discussion closely follows those of \Refs{Larkoski:2013eya,Larkoski:2014wba}, which in turn are based on the framework developed in \Refs{Dokshitzer:1998kz,Banfi:2004yd,Banfi:2004nk}.
Our primary goal here is to provide some additional clarification on technical points that may be less familiar to reader not as versed in the details of QCD resummation.
For a recent introduction to the general principles of final state resummation accessible to non-experts, see \Ref{Luisoni:2015xha}.

We begin with the collinear limit of the \ECF distribution, which is doubly divergent due to a collinear logarithm from the angular integral and a soft logarithm from the integral over the so-called splitting functions.
These splitting functions $p_i(z)$, which depend on the momentum fraction $z$ can be used to derived resummed distributions.
The leading order (LO) contribution is due to a single emission.
This can be simply modeled by integrating the splitting function against a delta function that enforces the 2-body momentum conservation as applied to \cref{eqn:e2beta}.
To this order, the differential distribution is
\begin{equation}
\label{eqn:sigmaLO_A}
  \frac{1}{\sigma} \dv{\sigma_i^\text{LO}}{\ECF}
    = \frac{\aS}{\pi} \int_0^{R_0} \frac{\dd{\theta}}{\theta} \int_0^1 \dd{z} 
       p_i(z)\, \delta\Biggl( z(1-z) \biggl(\frac{\theta}{R_0}\biggr)^\beta - \ECF \Biggr)\,,
\end{equation}
where $p_i(z)$ is the appropriate parton splitting function for a quark-initiated jet or a gluon-initiated jet, which are given by
\begin{align}
  p_q(z) & = P_{g \leftarrow q}(z) = C_F \frac{1+z^2}{1-z}, \notag\\
  p_g(z) & = \frac{1}{2} P_{g \leftarrow g}(z) + n_F P_{q \leftarrow g}(z) \notag\\
         & = C_A \biggl( \frac{z}{1-z} + \frac{1-z}{z} + z(1-z) \biggr) + n_F T_R \bigl( z^2+(1-z)^2 \bigr)\,.
\label{eqn:splittings_A}
\end{align}
For quark-initiated jets, only $P_{g \leftarrow q}$ is included, since the function $P_{q \leftarrow q}$ is not divergent in the soft limit and would effectively double count the jet core.
Likewise, for gluon-initiated jets, the factor of $\frac{1}{2}$ multiplying $P_{g \leftarrow g}$ accounts for a double counting that results from there being the two gluons emerging from a single gluon, while the factor of $n_F$ multiplying $P_{q \leftarrow g}$ provides the proper counting statistics for the gluon to split into $n_F$ different quark pairs.

In the limit where $\ECF \ll 1$, we can simplify $z(1-z) (\theta/R_0)^{\beta} \simeq z(\theta/R_0)^{\beta}$ by assuming $z \ll 1$.
It is then straightforward to evaluate \cref{eqn:sigmaLO_A}, which yields
\begin{align}
\label{eqn:sigmaLO_approx_A}
  \frac{\ECF}{\sigma} \dv{\sigma_i^\text{LO}}{\ECF} \simeq
    \frac{2\aS}{\pi} \frac{C_i}{\beta} 
      \Biggl(\ln\frac{1}{\ECF} + B_i + \ord\Bigl(\ECF\Bigr) \Biggr),
\end{align}
where $C_q = C_F = \frac{N_C^2-1}{2N_C}$ and $C_g = C_A = N_C$ are the color factors associated with the jet and $B_q = -\frac{3}{4}$ and $B_g = -\frac{11}{12} + \frac{n_F T_R}{3C_A}$ encode the subleading terms in the splitting functions and arise from hard collinear emissions. 
At LO, the cumulative distribution exhibits a characteristic double logarithm in the limit of small \ECF. Denoting the logarithm as $L \equiv \ln\frac{1}{\ECF}$, one finds
\begin{align}
  \Sigma_i^\text{LO} &\equiv
    \int_0^{\ECF} \dd{x} \frac{1}{\sigma} \dv{\sigma_i^\text{LO}}{x} =
    1 - \int_{\ECF}^{e_{2,\text{max}}^{(\beta)}}
          \dd{x} \frac{1}{\sigma} \dv{\sigma_i^\text{LO}}{x}\notag\\
          & =
    1 - \frac{\aS}{\pi} \frac{C_i}{\beta} \Bigl( L^2 + 2B_i L + \ord(1) \Bigr).
\label{eqn:cumLO_approx_A}
\end{align}
Note that the first integral is divergent, since we have not accounted for virtual corrections. 
However, we can sidestep this issue by assuming that the probability to emit anywhere is finite.
Instead of computing the missing $\ord(\aS)$ corrections to the total rate, we instead invoke unitary to write the integral in the second finite form which implicitly includes the virtual corrections.

Due to presence of the logarithm in \cref{eqn:cumLO_approx_A}, perturbative control over the differential distribution is lost for small values of \ECF.
Particles with different color charges are going to give qualitatively different behavior in precisely this limit, and so it is necessary to resum the resulting logarithms to all orders to explore how the distributions differ.
To leading-log (LL) accuracy, one can consider the emission of $n$ collinear partons within the jet as independent, with the scale of the (one-loop) coupling for each splitting $m$ chosen at the relative transverse momentum scale $\kappa_m = z_m \theta_m p_{T_J}$.
Virtual corrections do not change the kinematics, so they will contribute to the distribution for any value of the observable, whereas real emissions will only contribute if the kinematic configuration is such that the emission angle is smaller than the jet radius.
At LO, virtual corrections only yield a divergent correction to the tree-level value of $\ECF = 0$.
Thus, to LL accuracy, the resummed cumulative distribution can be computed by simply summing over all emissions off the initial parton while treating them as uncorrelated. 
In the small $z$ limit, and taking the second form of the integral in \cref{eqn:cumLO_approx_A} to work with finite quantities, the resummed cumulative distribution is given by
\begin{align}
  \Sigma_i^\text{LL}
    &= \sum_{n=0}^\infty \frac{1}{n!} \prod_{m=1}^n \int_0^{R_0} \frac{\dd{\theta_m}}{\theta_m} 
      \int_0^1 \dd{z_m} p_i(z_m) \frac{\aS(\kappa_m)}{\pi} \Biggl(
      \Theta\Biggl( \ECF - z_m \biggl(\frac{\theta_m}{R_0}\biggr)^\beta \Biggr) - 1 \Biggr) \notag\\
    &= \sum_{n=0}^\infty \frac{(-1)^n}{n!} \prod_{m=1}^n \int_0^{R_0} \frac{\dd{\theta_m}}{\theta_m}
      \int_0^1 \dd{z_m} p_i(z_m) \frac{\aS(\kappa_m)}{\pi}
      \Theta\Biggl( z_m\biggl(\frac{\theta_m}{R_0}\biggr)^\beta - \ECF \Biggr)\,,
\label{eqn:cumLL_A}
\end{align}
where the second line sums over virtual emissions, which have the same matrix element as real emissions by unitarity (modulo a sign difference)~\cite{Banfi:2004yd,Banfi:2004nk}.
The series above is readily resummed into a single term, correct to double logarithmic accuracy:
\begin{align}
  \Sigma_i^\text{LL} &= e^{-R_i}\,,\notag\\
  R_i &= \int_0^{R_0} \frac{\dd{\theta}}{\theta} \int_0^1 \dd{z} p_i(z) \frac{\aS(\kappa)}{\pi} \,
  \Theta\Biggl( z\biggl(\frac{\theta}{R_0}\biggr)^\beta - \ECF \Biggr)\,.
\label{eqn:radiatorLL_A}
\end{align}
The function $R_i$ is called the radiator for the jet, and it captures the Sudakov double logarithms associated with the IR divergences that result from soft or collinear emissions from the hard parton.
In the fixed coupling approximation, the radiator has the form
\begin{equation}
\label{eqn:radiatorFO_A}
  R_i \simeq \frac{\aS}{\pi} \frac{C_i}{\beta} \Bigl( L^2 + 2B_i L + \ord(1) \Bigr)\,,
\end{equation}
so that expanding $\Sigma^\text{LL}$ to leading order in the radiator recovers the LO behavior in \cref{eqn:cumLO_approx_A}.

At NLL order a number of new effects appear: multiple emissions, the two-loop running coupling, and non-global logarithms that arise from out-of-jet-emissions falling within the cone. 
The resummed cumulative distribution can be improved to single logarithmic accuracy by explicitly summing over uncorrelated emissions:\footnote{This formula ignores the effects of non-global logarithms, which must be separately implemented to achieve true NLL accuracy.}
\begin{equation}
\label{eqn:cumNLL_A}
\begin{aligned}
  \Sigma_i^\text{NLL} &= \sum_{n=0}^\infty \frac{1}{n!} \prod_{m=1}^n \int_0^{R_{0}} \frac{\dd{\theta_m}}{\theta_m} \int_0^1 \dd{z_m} p_i(z_m) \frac{\aS(\kappa_m)}{\pi}\s\Theta(\theta_{m-1}-\theta_m)\\
  &\times\Theta\Biggl(\ECF - \sum_{m=1}^n z_m \biggl(\frac{\theta_m}{R_0}\biggr)^\beta \Biggr)
  e^{-\int_0^{R_0}\frac{\dd{\theta}}{\theta} \int_0^1 \dd{z}\s p_i(z) \frac{\aS(\kappa)}{\pi}}\,.
\end{aligned}
\end{equation}
The angular ordering condition comes from the fact that when inserting an eikonal emission factor $\sum_i \boldsymbol{T}_i k_i \cdot \epsilon/ k_i \cdot q$ into an existing matrix element $\mathcal{M}$, the squared matrix element picks up a kinematic factor of
\begin{equation}
  |\mathcal{M}|^2 \sim \sum_{i < j} W_{ij}\,, \quad \text{where} \quad
  W_{ij} = \frac{1 - \cos\theta_{ij}}{(1-\cos\theta_{iq})(1-\cos\theta_{jq})}\,.
\end{equation}
Each such term can be rewritten as $W_{ij} = W_{ij}^{(i)} + W_{ij}^{(j)}$, where
\begin{equation}
  W_{ij}^{(i)}= \frac{1}{2} \Biggl( W_{ij} + \frac{1}{1 - \cos\theta_{iq}}
                - \frac{1}{1 + \cos\theta_{jq}}\Biggr).
\end{equation}
The benefit of this rewriting is that every such term satisfies an angular ordering property,
\begin{equation}
  \int_0^{2\pi} \frac{\dd{\phi_{iq}}}{2\pi}\, W_{ij}^{(i)} =
    \begin{cases}
      \frac{1}{1 - \cos\theta_{iq}} & \theta_{iq} < \theta_{ij}\\
      0 & \text{otherwise}
    \end{cases}\,,
\end{equation}
such that the soft limits are correctly reproduced through the treatment of collinear divergences and angular ordering together.
The resulting expression can be evaluated in Laplace space, where the convolution of the splitting function, running coupling, and the $\Theta$ functions become a summable product, yielding~\cite{Banfi:2004yd} 
\begin{align}
  \Sigma_i^\text{NLL} &= \int\frac{\dd{\nu}}{2\pi i\nu}\,e^{\ECF} e^{-R_i}\,,\notag \\
  R_i &= \int_0^{R_0} \frac{\dd{\theta}}{\theta} \int_0^1 \dd{z} p_i(z) \frac{\aS(\kappa)}{\pi} \biggl(1-e^{-\nu z\bigl(\frac{\theta}{R_0}\bigr)^{\beta}}\biggr)\,,
\label{eqn:radiatorNLL_A}
\end{align}
where $R_i$ here is the Laplace space version of the expression in \cref{eqn:radiatorLL_A}.

Logarithmic accuracy in $\nu$ tracks the logarithmic accuracy in \ECF, since they are Laplace conjugates of each other. 
Therefore, to derive the NLL cumulative distribution, one must compute the radiator to single logarithmic accuracy in $\nu$. 
Expanding about $\nu^{-1} = \ECF$ gives
\begin{align}
  \Sigma_i^\text{NLL} &= N\, \frac{e^{-\gamma_E R'_i}}{\Gamma(1+R'_i)}\, e^{-R_i}\,,\notag\\
   R'_i &\equiv \dv{R_i}{L}\,,
\label{eqn:cumNLL_solved_A}
\end{align}
where $N = 1 + \ord(\aS)$ is a matching coefficient that can be determined by comparing with the fixed-order cumulative distribution, $\gamma_E$ is the Euler--Mascheroni constant, and the radiator $R_i$ is given in \cref{eqn:radiatorLL_A}.

Note that improving predictability to NLL order requires matching the resummed calculation to the fixed-order distribution.
To this end, we implement the Log-$R$ matching scheme~\cite{Catani:1992ua} by first considering the LO cumulative distribution, \ie, the properly integrated form of \cref{eqn:sigmaLO_A},
\begin{equation}
\label{eqn:cumLO_A}
  \ln\Sigma_i^\text{LO} =
    \frac{\aS}{\pi} \int_0^{R_0} \frac{\dd{\theta}}{\theta}\int_0^1 \dd{z} p_i(z) 
    \s\Theta\Biggl( z\biggl(\frac{\theta}{R_0}\biggr)^\beta - \ECF \Biggr) =
      -\frac{\aS}{\pi} R_{1,i}\,,
\end{equation}
where
\begin{align}
R_{1,q} = \frac{C_F}{\beta}\Biggl( & -4\Li_2\biggl(\frac{1+u}{2}\biggr) + 3u + \ln^2(1-u) - 2\ln(1+u)\ln(1-u)\notag\\[-5pt]
& + \bigl(4\ln2-\ln(1+u)\bigr) \ln(1+u) - 3\tanh^{-1}u + \frac{\pi^2}{3} - 2\ln^22\Biggr)\,,\notag\\[8pt]
R_{1,g} = \frac{C_A}{\beta}\Biggl( & -4\Li_2\biggl(\frac{1+u}{2}\biggr) + \biggl(\frac{67}{18} - \frac{2}{9}C_1^{(\beta)}\biggr)u - \frac{n_F}{C_A}\biggl(\frac{13}{18} - \frac{2}{9} C_1^{(\beta)}\biggr)u\notag\\
& +\ln^2(1-u) - 2\ln(1+u)\ln(1-u) + \bigl(4\ln2 - \ln(1+u)\bigr)\ln(1+u)\notag\\
&  -\biggl(\frac{11}{3} - \frac{2n_F}{3C_A}\biggr) \tanh^{-1}u + \frac{\pi^2}{3} - 2\ln^2 2 \Biggr)\,,
\label{eqn:FOcorrection_A}
\end{align}
and $u \equiv \sqrt{1 - \ECF}$.
Here $u$ takes values between $u = \sqrt{1 - e_{2,\max}} = \sqrt{1 - \frac{1}{4}R_0^{\beta}}$ and 1.
With the Log-$R$ matching scheme, it is straightforward to match the resummed and fixed-order results,
\begin{equation}
\label{eqn:cumNLL_matched_A}
\Sigma_i^\text{NLL} = N \frac{e^{-\gamma_E R_i'}}{\Gamma(1+R_i')}
\exp({-R_i}) \,\exp({-\frac{\aS}{\pi} (R_{1,i} - G_{2,i} L^2 - G_{1,i} L)})\,,
\end{equation}
where $G_{2,i} L^2$ and $G_{1,i} L$ are the logarithms appearing in the fixed-order expression which must to be subtracted from $R_{1,i}$ to avoid double counting the resummed logarithms.
From \cref{eqn:cumLO_approx_A}, these logarithms are explicitly
\begin{equation}
\label{eqn:logmatching_A}
G_{2,i} L^2 + G_{1,i} L = \frac{C_i}{\beta} \bigl( L^2 + 2B_i L \bigr).
\end{equation}

Using this analytic form in \cref{eqn:cumNLL_matched_A} requires evaluating the radiator $R_{i}$, which is given in \cref{eqn:radiatorLL_A}.
An analytic evaluation of $R_i$ is possible, although challenging, \eg~see~\Ref{Marzani:2017mva}.
The calculation of the resulting efficiencies at NLL due to a cut on \ECF requires evaluating the gauge coupling \aS at two-loop order using the CMW scheme~\cite{Catani:1990rr}, such that efficiencies still need to be computed numerically.
Another issue is related to \aS becoming non-perturbative as the integral is evaluated at low enough scales.
Following the procedure in \Ref{Larkoski:2014wba}, the coupling is only run at one-loop order and is frozen at the non-perturbative scale $\mu_\text{NP} \equiv 7\Lambda$.
These choices result in a closed-form solution for $R_{i}$ while limiting its logarithmic accuracy, so \cref{eqn:cumNLL_matched_A} provides a modified leading logarithmic (MLL) resummed cumulative distribution with FO corrections.
All analytic distributions presented above are to LL or MLL+FO accuracy, but strictly not accurate to full NLL order.

The prescription of freezing the coupling at the non-perturbative scale $\mu_\text{NP} \equiv 7\Lambda$ leads to the explicit form
\begin{equation}
\label{eqn:couplingNP_A}
  \aS(\kappa) = \aS^\text{(1L)}(\kappa) \Theta(\kappa - \mu_\text{NP})
                + \aS^\text{(1L)}(\mu_\text{NP}) \Theta(\mu_\text{NP}-\kappa)\,,
\end{equation}
where $\aS^\text{(1L)}(\kappa)$ is the standard one-loop order expression for the coupling,
\begin{equation}
\label{eqn:coupling1loop_A}
  \aS^\text{(1L)}(\kappa) = \frac{1}{2\beta_0\ln\bigl(\frac{\kappa}{\Lambda}\bigr)}\,.
\end{equation}
For brevity, we introduce the function $F(x) = x\ln x$ and define the following variables:
\begin{equation}
\label{eqn:analyticvariables_A}
\begin{aligned}
  L = \ln\frac{1}{\ECF} & \qquad \qquad \qquad L_\mu = \ln\frac{1}{\mu}\\
  \lambda = 2\aS\beta_0 L & \qquad \qquad \qquad \lambda_\mu = 2\aS \beta_0 L_\mu
\end{aligned},
\end{equation}
where $\mu = \frac{\mu_\text{NP}}{p_T R_0}$ is the relevant scale associated with the non-perturbative transition.

Finally, we will write down the explicit expressions for the radiator functions that are used here.  
Their form depends on the choice of the angular dependence $\beta$.
For $\beta > 1$,
\begin{align}
\label{eqn:radiator_g1_A}
R_i^{(\beta > 1)} =
  \begin{cases}
    \frac{C_i}{2\pi\aS\beta_0^2} \biggl(\frac{F(1-\lambda)}{\beta-1}
      - \frac{\beta F(1-\frac{1}{\beta}\lambda)}{\beta-1}
      - 2\aS\beta_0 B_i\ln\Bigl(1-\frac{1}{\beta}\lambda\Bigr)\biggr)
    & \ECF > \mu \\[20pt]
    \frac{C_i}{2\pi\aS\beta_0^2} \bigg(\frac{F(1-\lambda_\mu)}{\beta-1}
      - \frac{\beta F(1-\frac{1}{\beta}\lambda)}{\beta-1}
      - \frac{1 + \ln(1-\lambda_\mu)}{\beta-1} (\lambda-\lambda_\mu) \\
      \hspace{1.65cm} {} - 2\aS\beta_0 B_i \ln\Bigl(1-\frac{1}{\beta}\lambda\Bigr)\bigg)
                      + \frac{C_i\aS(\mu_\text{NP})}{\pi} \frac{(L-L_\mu)^2}{\beta-1}
    & \mu \ge \ECF > \mu^\beta \\[20pt]
    \frac{C_i}{2\pi\aS\beta_0^2} \Big(-F(1-\lambda_\mu) 
      - \bigl(1+\ln(1-\lambda_\mu)\bigr)\lambda_\mu \\
      \hspace{1.65cm} {} - 2\aS\beta_0B_i\ln\bigl(1-\lambda_\mu\bigr)\Big) \\
      \hspace{1.75cm} {} + \frac{C_i\aS(\mu_\text{NP})}{\pi}
        \Bigl((\beta-1) L_\mu^2 + \frac{L-\beta L_\mu}{\beta}(L+\beta L_\mu + 2B_i)\Bigr)
    & \mu^\beta \ge \ECF
  \end{cases}\,,
\notag\\[-20pt] 
\notag\\
\end{align}
while for $\beta < 1$,
\begin{align}
\label{eqn:radiator_l1_A}
R_i^{(\beta<1)} =
  \begin{cases}
    \frac{C_i}{2\pi\aS\beta_0^2} \biggl(\frac{F(1-\lambda)}{\beta-1}
      - \frac{\beta F(1-\frac{1}{\beta}\lambda)}{\beta-1}
      - 2\aS\beta_0B_i\ln\Bigl(1-\frac{1}{\beta}\lambda\Bigr)\biggr)
    &\hspace{25pt} \ECF > \mu \\[20pt]
    \frac{C_i}{2\pi\aS\beta_0^2} \biggl(\frac{F(1-\lambda)}{\beta-1}
      - \frac{\beta F(1-\lambda_\mu)}{\beta-1}
      - \frac{1+\ln(1-\lambda_\mu)}{\beta-1} \bigl(\lambda-\beta\lambda_\mu\bigr) \\
      \hspace{1.65cm} {} - 2\aS\beta_0 B_i \ln\Bigl(1-\lambda_\mu\Bigr)\biggr) \\
      \hspace{1.85cm} {} + \frac{C_i\aS(\mu_\text{NP})}{\pi} \frac{L-\beta L_\mu}{\beta}
                          \frac{L-\beta L_\mu + 2(1-\beta)B_i}{1-\beta}
    &\hspace{25pt} \mu \ge \ECF > \mu^\beta \\[20pt]
    \frac{C_i}{2\pi\aS\beta_0^2} \Bigl(-F(1-\lambda_\mu)
      - \bigl(1+\ln(1-\lambda_\mu)\bigr)\lambda_\mu \\
      \hspace{1.65cm} {} - 2\aS\beta_0 B_i \ln\bigl(1-\lambda_\mu\bigr)\Bigr) \\
      \hspace{1.85cm} {} + \frac{C_i\aS(\mu_\text{NP})}{\pi}
                             \frac{L(L+2B_i)-\beta L_\mu(L_\mu+2B_i)}{\beta}
    &\hspace{25pt} \mu^{\beta} \ge \ECF
  \end{cases}\,,
  \notag\\[-20pt] 
\notag\\
\end{align}
Finally, in the limit $\beta \to 1$, \cref{eqn:radiator_g1_A} and \cref{eqn:radiator_l1_A} match:
\begin{align}
\label{eqn:radiator_e1_A}
  R_i^{(\beta=1)} = \begin{cases}
    \frac{C_i}{2\pi\aS\beta_0^2} \Bigl(-F(1-\lambda)
      - \bigl(1+\ln(1-\lambda)\bigr) \lambda - 2\aS\beta_0 B_i\ln(1-\lambda)\Bigr)
    & \ECF^{(1)} > \mu \\[20pt]
    \frac{C_i}{2\pi\aS\beta_0^2} \Bigl(-F(1-\lambda_\mu)
      - \bigl(1+\ln(1-\lambda_\mu)\bigr) \lambda_\mu
      - 2\aS\beta_0 B_i\ln(1-\lambda_\mu)\Bigr)  \\
      \hspace{1.65cm} {} +\frac{C_i\aS(\mu_\text{NP}}{\pi}(L - L_\mu) (L + L_\mu + 2B_i)
    & \mu \ge \ECF^{(1)}
\end{cases}\,.
\notag\\[-20pt] 
\notag\\
\end{align}

\bibliographystyle{utphys}
\bibliography{DarkShowerSystematics}

\end{document}